\documentclass{SciPost}

\hypersetup{
    colorlinks,
    linkcolor={red!50!black},
    citecolor={blue!50!black},
    urlcolor={blue!80!black}
}

\usepackage[bitstream-charter]{mathdesign}
\DeclareSymbolFont{usualmathcal}{OMS}{cmsy}{m}{n}
\DeclareSymbolFontAlphabet{\mathcal}{usualmathcal}

\fancypagestyle{SPstyle}{
\fancyhf{}
\lhead{\colorbox{scipostblue}{\bf \color{white} ~SciPost Physics }}
\rhead{{\bf \color{scipostdeepblue} ~Submission }}

\fancyfoot[C]{\textbf{\thepage}}
}

\renewcommand{\l}{\left(}
\renewcommand{\r}{\right)}
\newcommand{\lb}{\left[}
\newcommand{\rb}{\right]}
\newcommand{\lcb}{\left\{ }
\newcommand{\rcb}{\right\} }
\newcommand{\bra}{\left<}
\newcommand{\ket}{\right>}
\newcommand{\lv}{\left|}
\newcommand{\rv}{\right|}

\newcommand{\mc}[1]{\mathcal{\mathrel{#1}}}

\newcommand{\pI}{\mathbb{I}}
\newcommand{\pX}{\mathbf{X}}

\newcommand{\pZ}{\mathbf{Z}}

\newcommand{\npZ}{\hat{\mathbf{Z}}}

\renewcommand{\log}[1]{\, {\rm Log} \left[ \mathrel{#1} \right] }
\newcommand{\titlemath}[1]{\texorpdfstring{$\mathrel{#1}$}{TEXT}}
\renewcommand{\=}[1]{$\mathrel{#1}$}
\newcommand{\tr}[1]{{\rm Tr}\lb\mathrel{#1}\rb}

\newcommand{\svn}[1]{S_{\rm vN}^{\rm \mathrel{#1}}}
\newcommand{\mbs}[1]{\boldsymbol{#1}}
\renewcommand{\mod}[1]{\, {\rm mod} \; #1}

\begin{document}

\pagestyle{SPstyle}

\begin{center}{\Large \textbf{\color{scipostdeepblue}{
Long-Time Limits of Local Operator Entanglement in Interacting Integrable Models\\
}}}\end{center}

\begin{center}\textbf{
J. Alexander Jacoby\textsuperscript{1$\star$} 
and Sarang Gopalakrishnan\textsuperscript{2$\dagger$}
}\end{center}

\begin{center}
{\bf 1} Department of Physics, Princeton University, Princeton, New Jersey 08544, USA
\\
{\bf 2} Department of Electrical and Computer Engineering, Princeton University, Princeton NJ 08544, USA
\\[\baselineskip]
$\star$ \href{mailto:ajacoby@princeton.edu}{\small ajacoby@princeton.edu}\,,\quad
$\dagger$ \href{mailto:sgopalakrishnan@princeton.edu}{\small sgopalakrishnan@princeton.edu}
\end{center}

\section*{\color{scipostdeepblue}{Abstract}}
\textbf{\boldmath{%
    We explore the long-time behavior of Local Operator Entanglement entropy (LOE) in finite-size interacting integrable systems. For certain operators in the Rule 54 automaton, we prove that the LOE saturates to a value that is at most logarithmic in system size. This bound extends previous work [PRL \textbf{122}, 250603; Commun. Math. Phys. 371, 651-688] showing LOE grows logarithmically in the early time regime, $t\ll L$, to the late time regime, $t\gg L $.
    However, the late-time logarithmic bound relies on a feature of Rule 54 that does not generalize to other interacting integrable systems: namely, that there are only two types of quasiparticles, and therefore only two possible values of the phase shift between quasiparticles. We present a heuristic argument, supported by numerical evidence, that for generic interacting integrable systems (such as the Heisenberg spin chain) the saturated value of the LOE is volume-law in system size. 
}}

\vspace{\baselineskip}



\vspace{10pt}
\noindent\rule{\textwidth}{1pt}
\tableofcontents
\noindent\rule{\textwidth}{1pt}
\vspace{10pt}


\section{Introduction}
\subsection{Background}
Finding metrics that sharply distinguish between quantum chaotic and integrable dynamics has proved challenging. Classically, the Lyapunov spectrum (which quantifies the rate at which nearby trajectories diverge) serves as a diagnostic of chaos \cite{Pesin_76,Pesin_77}. However, the notion of trajectories is not well-defined in quantum systems, outside of certain semiclassical limits. In these semiclassical limits, the Lyapunov exponents can be computed from the growth rate of the out-of-time-order correlator (OTOC)  \cite{Larkin,Roberts_2015,Aleiner_2016,Maldacena_2016,Roberts_2017,Roberts_2018}. OTOCs have been measured in experiments on intermediate-scale quantum devices \cite{Swingle_2016,Li_2017, G_rttner_2017}. 

Heuristically, the OTOC describes how the footprint of an initially local operator spreads under time evolution in the Heisenberg picture, as it becomes increasingly nonlocal (and therefore unobservable in practice). Indeed, the average of the OTOC over all operators supported in a spatial region can be rigorously related to information-theoretic measures of scrambling \cite{dowling2023scrambling,dowling2023operational}. It is unclear, however, to what extent these information-theoretic measures distinguish between chaotic systems with a few conservation laws and integrable systems with extensively many conservation laws. More generally, one is interested in the dynamics of \emph{specific}, initially simple operators, and for these there is no direct relation between the OTOC and information scrambling. Instead, explicit calculations of generic OTOCs in integrable systems show features that are qualitatively (and in some cases quantitatively) very similar to the behavior expected in chaotic systems or random unitary circuits \cite{Lin_Int_OTOC, GopalakrishnanR54-1, Nahum_2018,Keyserlingk_OpRUCs}. The essential challenge is that although the many-body eigenstates of integrable systems are labeled by quasiparticle occupation numbers, the action of local operators on the quasiparticle states is highly nontrivial \cite{Essler_2024}.

An obvious drawback of the OTOC is that it probes the ``size'' of an operator, rather than its complexity. One possible scenario for interacting integrable dynamics is that operators spread while remaining in some sense simple: unitary Clifford circuits furnish an extreme example of this scenario, where a single-site Pauli operator evolves to a \emph{single} long string of Pauli operators. A metric that attempts to quantify the complexity of an operator is its operator-space entanglement entropy (which has recently been related to the non-Cliffordness of evolution \cite{Dowling_2025}). To compute this, one writes the operator $O: \mathcal{H} \to \mathcal{H}$ as a normalized state $|O\rangle \in \mathcal{H} \otimes \tilde{\mathcal{H}}$ in a doubled Hilbert space, and computes its entanglement entropy as one would for a state (we define this explicitly below). We will be concerned with operators that are initially local (as opposed to, say, the time evolution operator itself as in \cite{Zanardi,Zhou_2017}), and accordingly call this metric the ``local operator entanglement'' (LOE).

The growth of LOE was first considered for integrable systems that map onto free fermions \cite{Prosen_Znidaric, Prosen_Transverse_Ising,Prosen_XY} (see also \cite{Dubail_2017}). In these systems, the LOE grows as $\log t$ or slower; its saturated value for a subsystem of size $\ell$ scales as $\log \ell$. This behavior contrasts with that seen in generic chaotic systems, where the membrane picture \cite{Jonay,Kudler_Flam_2020,Bertini_2020_1} predicts linear growth in time and a saturation value that scales linearly in $\ell$, i.e., with a volume law. The intermediate case of interacting integrable systems was first discussed in Refs. \cite{Klobas_2019,Alba_2019, Alba_2021} (see also \cite{Alba_2024}) for the integrable cellular automaton known as Rule 54. These works established that the half-system LOE at early times (i.e., times much shorter than the system size $L$) scales as $S_O(t) \sim  \log t$. Similar bounds were conjectured to hold for interacting integrable systems in general, based on suggestive numerical evidence. Why this scaling should hold beyond the exactly solvable case of Rule 54, however, has remained unclear. 

Our main objective in the present work is to explore the late-time asymptotics of the half-system LOE in finite-size systems. We use periodic boundary conditions to ensure that integrability is preserved at all times. Taking the late-time limit in a finite-size system allows us to make simple arguments based on the dephasing between many-body eigenstates with different quasiparticle content. These arguments lead us to the following key conclusions, supported by numerics: (i)~in Rule 54, the LOE saturates to a value that is at most logarithmic in system size---a result we can prove for this simple model; (ii)~in the Heisenberg spin chain, our representative example of a generic interacting integrable model, the (von Neumann) LOE instead seems to reach a volume law, a result that our numerics and a simple picture based on dephasing both suggest; and (iii)~the R\'enyi entropies for $\alpha > 1$ saturate at values logarithmic in system size in integrable models, as a trivial consequence of the many conservation laws. After discussing these results, we comment on how they might relate to previous conjectures concerning the finite-time growth of LOE.

\subsection{Defining local operator entanglement}

We start with an operator $O: \mathcal{H} \to \mathcal{H}$, which can be written in a reference basis (we choose the computational basis) as $O = \sum_{xy} O_{xy} \,  |x\rangle \langle y|$. To convert this into a state, one flips the bras to kets, giving the state $|O\rangle \propto \sum_{xy} O_{xy} |x\rangle \otimes |\tilde{y}\rangle \in \mathcal{H} \otimes \tilde{\mathcal{H}}$ (up to normalization). We normalize $|O\rangle$ so that $\langle O | O\rangle = 1$. We then trace out all sites in the doubled Hilbert space that are outside the region $A$ of interest---we denote the complement of $A$ as $\bar A$. This gives the super-density matrix $\rho_O^{(A)} \equiv \mathrm{Tr}_{\bar A} (|O\rangle \langle O|)$. The operator entanglement entropies are the R\'enyi entropies of this super-density matrix, specifically:
\begin{equation}\label{loedef}
S_\alpha(|O\rangle; A) \equiv (1-\alpha)^{-1} \log{\mathrm{Tr}_A\left\{\mathrm{Tr}_{\bar A}(|O\rangle \langle O|)^\alpha\right\}}.
\end{equation}

The limit $\alpha \to 0$ is the logarithm of the minimum bond dimension required to represent $O$ as a matrix-product operator, and thus has a direct meaning in terms of computational complexity. We will primarily focus on the $\alpha \to 1$ limit, the von Neumann operator entanglement, but also comment on other values of $\alpha$. Unless otherwise stated, numerical results for LOE will be reported in units of log base $4$ (representing the natural unit for the doubled operator Hilbert space), which are equal to two bits. We will also often use the LOE density in these units, which is maximal (\textit{i.e.}, unity) at two bits per site.

\section{Rule 54 Cellular Automaton}

\label{sect:def_rule_54}
\subsection{Definition}
The Rule 54 Cellular Automaton (CA) was introduced as a model of integrable dynamics in \cite{Bobenko}. It is not Bethe-ansatz solvable in the customary sense, but can be regarded (like most integrable CAs) as a nondispersive limit of a family of Bethe-ansatz solvable integrable models~\cite{Friedman_2019}. Nonetheless, Rule 54 exhibits the essential features of interacting integrability in $1+1$ dimensions: stable, interacting chiral quasiparticle excitations.  In the automaton limit, these quasiparticles are nondispersive: they come in two flavors, left- and right-moving. Scattering events between opposite chirality quasiparticles impart a uniform time delay.

\begin{figure}[h]
    \centering
    \includegraphics[width = 0.75\linewidth]{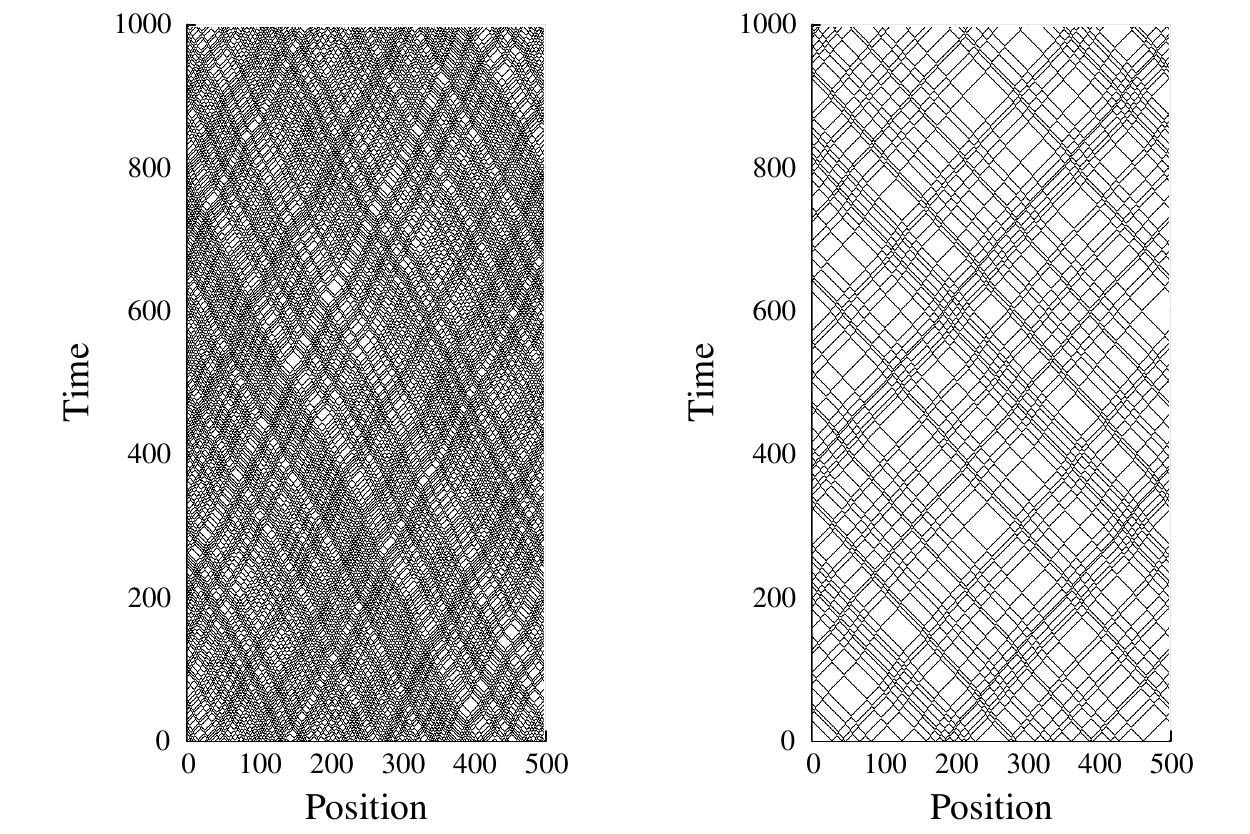}
    \caption{(Left) Rule 54 dynamics on a generic quasiparticle density ($n_{L}\approx n_{R} \approx L/4$) initial state. (Right) Rule 54 dynamics on a low quasiparticle density initial state.}
    \label{fig:R54}
\end{figure}

Computational basis states, or bitstrings, are evolved with a three-site update,
\begin{eqnarray}
    U_{n} &=& \lv 101 \ket\bra 111 \rv + \lv 100 \ket\bra 110 \rv + \lv 111 \ket\bra 101 \rv \nonumber  \\
    && + \lv 110 \ket\bra 100 \rv + \lv 001 \ket\bra 011 \rv + \lv 010 \ket\bra 010 \rv \nonumber \\
    &&+ \lv 011 \ket\bra 001 \rv + \lv 000 \ket\bra 000 \rv,
    \label{eqn:r54_u_n}
\end{eqnarray}
which acts on sites $\lcb n-1,n,n+1\rcb$. We can then define a full Floquet pump ($\delta t = 2$) as
\begin{equation}
    U_{54} = \l \ \prod_{n \, {\rm Even}}^{N} U_{n}\r \l \ \prod_{n \, {\rm Odd}}^{N} U_{n}\r.
    \label{eqn:U_R54}
\end{equation}
A simple example of the dynamics of the Rule 54 model is shown in Fig. \ref{fig:R54}.
We count the full update of Eqn. (\ref{eqn:U_R54}) as two time steps and each qubit as one spatial site. This choice implies unit velocity of quasiparticles while retaining the clarity of the computational basis states for entanglement calculations, but at the cost of non-scattering quasiparticles occupying two adjacent sites (position with respect to even/odd lattice sites decides quasiparticle chirality). Additionally, time delays are two time units rather than one.

An essential feature of this model on a periodic chain is that a full Floquet cycle, $U_{54}$, acts as a permutation on computational basis states. We will call this permutation $\sigma$. $\sigma$ possesses a cycle decomposition: $\sigma = \bigoplus_{k}\sigma^{k}$. This decomposition furnishes the dynamics with orbits of computational basis states spanning $\mc{H}_{k}$, where $\bigoplus_{k}\mc{H}_{k} = \mc{H}$. An orbit has length (recurrence time) $C_{k} = {\rm dim}\l \mc{H}_{k}\r$ and consists of computational basis states $\lv\phi^{k,\, p}\ket$ (with $p$ running from $1$ to $C_{k}$) satisfying $U_{ 54} \lv \phi^{k,p}\ket =\lv \phi^{k,\,\l p+1\r_{C_{k}} }\ket$, where we introduce the shorthand $\l\star\r_{C_{k}} = 1+  \l \star -1\r \mod{C_{k}}$. The eigenstates can be constructed as
\begin{equation}    \label{eqn:R54_eigstates}
    \lv E^{k,n}\ket = \sum_{p=1}^{C_{k}} e^{2\pi i np/C_{k}} \lv  \phi^{k, \, p }\ket.
\end{equation}
The eigenvalues of $U_{\rm 54}$ are thus $\l n \r_{C_{k}}$ multiples of the $C_{k}{\rm th}$ root of unity. Typical cycle length scales as $C_{k}\sim L^2$ \cite{GopalakrishnanR54-1}. As a result the eigenstates of Rule 54 have sub-volume-law von Neumann entanglement entropy,
\begin{equation}
    S^{\rm A }_{\rm vN} \sim 
    \begin{cases}
        \lv A\rv & \lv A \rv < 2\log{L} \\
        O\l 2\log{\lv L \rv}\r & \lv A \rv \gtrsim 2\log{L} 
    \end{cases}
\end{equation}
in nats for a bipartition of size $A$. However, Rule 54 is expected to exhibit ballistic von Neumann (though \textit{not} higher R\'enyi \cite{Bertini_2022}) entanglement entropy growth after a global quench from a generic product state following exactly the quasiparticle picture proposed by Calabrese and Cardy \cite{Calabrese_SVN_INT}.

\subsection{Rule 54 Conserved Quantities}
\label{sect:rule54_symmetry}
The Rule 54 chain has two types of conserved quantities: quasiparticle number and Asymptotic Spacings (AS). The AS refer to the distance between two quasiparticles of the same chirality in the absence of quasiparticles of the opposite chirality. This can be conceptualized on periodic boundary conditions by cutting the chain in an empty region. The AS can then be measured after the left and right movers separate out from each other. Given all AS of a computational basis state the recurrence time can always be calculated, though the converse is not true. Indeed, exponentially many (in $L$) distinct AS may possess the same (typical) recurrence time.

Quasiparticle number sectors consist of the subspaces with a fixed and distinct number of right and left moving quasiparticles, denoted $\vec{N} = \lb N_{L}, N_{R}\rb$. The number sectors on periodic boundary conditions can be counted:
\begin{equation}
    \#\l L \r = 
    \begin{cases}
       \frac{1}{8}L^{2} + \frac{1}{2}L + 1 & \  L \mod{4} = 0 \\
       \frac{1}{8} L^{2} + \frac{1}{2} L + \frac{1}{2} & \ L\mod{4} = 2
    \end{cases}
    .
    \label{eqn:sector_count}
\end{equation}
The above result is exact and accounts for correlations between the number of left and right movers. The average filling is $N_{L/R} \approx L/4$ and the maximum filling is $N_{L}=N_{R} = L/2 $. Sectors with the same quasiparticle occupations have approximately commensurate recurrence times; if $N^{\l k \r}_{L/R} = N^{\l l \r}_{L/R}$, then $C_{k}$ and $C_{l}$ typically differ by at most an $O\l 1 \r$ factor. We can perform a heuristic calculation which leverages this fact to simplify the problem of long-time LOE in the Rule 54 chain before we derive more rigorous bounds for diagonal operators along the same lines in Section \ref{sect:diag_analytics}.

Rule 54 maps diagonal operators to diagonal operators, so we can treat the diagonal elements as the states (in the sense that $\pI \to \lv + X\ket$ and $\pZ \to \lv -X \ket$ in the reduced Hilbert space). We can write the operator state for $\pZ_{x}\l t \r$ as
\begin{eqnarray}
    \lv \pZ_{x}\l t \r \ket &=& \frac{-1}{\sqrt{2^L}}\sum_{k,p}e^{i\pi\phi_{x}^{k,\, p }}\lv \phi^{k,\, \l p+t\r_{C_{k}}}\ket \nonumber  \\
    &=&\frac{-1}{\sqrt{2^L}}\sum_{k,p}e^{i\pi\phi_{x}^{k,\, \l p-t \r_{C_{k}}}}\lv \phi^{k,\, p}\ket ,
\end{eqnarray}
where we emphasize the shorthand notation $\l n \r_{C_{k}} =\l n-1 \r \, \mod{C_{k}} + 1 $ and introduce $\phi^{k,p}_{x}$ as the $x$th entry of \={\phi^{k, p}} (recall these are the bitstrings in cycle $k$).

Then, we can transform to the energy eigenbasis using Eqn. (\ref{eqn:R54_eigstates}) to get
\begin{eqnarray}\label{eqn:energy_matrix_els}
    &&   \bra E^{k,\, n} \rv \rho_{\pZ_{x}}\l t \r\lv E^{l,\, m} \ket \\
     &&=\sum_{p,q}\exp{\lb 2\pi i \l \frac{mq}{C_{l}}-\frac{np}{C_{k}}+ \frac{\phi_{x}^{k,\, \l p - t\r_{C_{k}}}- \phi_{x}^{l,\, \l q - t\r_{C_{l}}}}{2}\r\rb} . \nonumber
\end{eqnarray}

Define $N_{L/R}$ to be the quasiparticle occupation of $\lv E^{l, m } \ket$ and $\tilde{N}_{L/R}$ to be the quasiparticle occupation of $\bra E^{k,n} \rv$. Then for $N_{L/R}\neq \tilde{N}_{L/R}$ the sign of the summand in Eqn. (\ref{eqn:energy_matrix_els}), given by $\exp\lb i\pi \l \phi_{x}^{k,\, \l p - t\r_{C_{k}}}- \phi_{x}^{l,\, \l q - t\r_{C_{l}}}\r\rb $, is pseudo-random since $C_{k}$ and $C_{l}$ are incommensurate if $N_{L/R}$ and $\tilde{N}_{L/R}$ are unequal. When the operator is marginalized, we sum contributions from different global number sectors and the coherences resulting from off-diagonal elements in this basis are therefore suppressed as a sum of random phases. That is, we expect \textit{dephasing} to occur in the number-sector basis.

\subsection{Diagonal Operator Bound}
\label{sect:diag_analytics}

Above, we argued that the super-density matrix $\rho_{\pZ_{x}}\l t \r$ should look approximately block-diagonal in the quasiparticle number eigenbasis at late times. We cannot rigorously show that dephasing happens. Instead, we upper-bound the LOE by
showing that $\lv \pZ_{x}\ket$ can be written as a sum of at most $O\l L^{8}\r$ terms which factorize between $A$ and $\bar{A}$ with contributions naturally organized by the number sectors in which coherent evolution occurs. We start in the diagonal operator state mapping (\textit{i.e.},  neglecting the dual-space for diagonal operators). Let $\lcb\pi_{\vec{N}}\rcb$ (where $\vec{N} = \lb N_{L},N_{R} \rb$) be the orthogonal projectors onto the subspace of $\mc{H}$ with $N_{L/R}$ left/right movers; these operators are discussed further in Appendix \ref{app:number_projectors}. We can then write
\begin{equation}
    \lv  \pZ_{x}\l t \r\ket = \sum_{\vec{N}} \pi_{\vec{N}} \lv \pZ_{x}\l t \r \ket
    \label{eqn:number_resolution}
\end{equation}
identically since $\sum_{\vec{N}}\pi_{\vec{N}} = \mathbb{I} $. There are at most $O\l L^{2}\r $ terms in this sum per Eqn.~\ref{eqn:sector_count}.

We now turn to the bond dimension of the wavefunction projected into each quasiparticle number sector. Our objective will be to show that this is also $O({\rm Poly}\l L\r )$ at all times. Our argument will rely on three main claims. First, the state corresponding to any local diagonal operator can be expressed as linear combination of $O(1)$ terms in the quasiparticle basis, where each term ``tags'' $O(1)$ quasiparticles near the operator insertion site. Second, in each quasiparticle number sector, any such initial states recur (up to a global translation) on a timescale $t \sim L$ that we will call the self-scattering time. At the self-scattering time, each left (right) moving quasiparticle has scattered off every right (left) moving quasiparticle exactly once. The self-scattering time is sharp because the model is dispersionless. Third, we will write the time-evolved operator state projected into a quasiparticle number sector, at any time before the self-scattering time, as a sum of $O(L^2)$ terms, each of which can be expressed as a matrix product state (MPS) with bond dimension $O(L^4)$. This gives us the bond dimension bound of $O \l L^{6} \r $ for each of the terms of Eqn.~\ref{eqn:number_resolution}; putting this together with the number of terms in Eqn.~\ref{eqn:number_resolution} we will arrive at the final bound
\begin{equation}
    S_{\rm vN} \leq S_{\l 0 \r} \lesssim 8 \, {\rm Log}_{2}\lb L \rb.
\end{equation}
This bound is expected to be very loose; it should be possible to reduce the prefactor, but we will not attempt this here as our main objective is just to show that the LOE will never exceed $O(\log L)$. 

\emph{Expanding the operator}.---As discussed above, we will think of a diagonal operator as a state $|O\rangle \in \mathcal{H}$. This state is a product of $\pi^{1}_{x} + \pi^{0}_{x} \to |\mathbb{I}_{x}\rangle$ everywhere except at the operator insertion, where $\pi_{x}^{0/1}$ is the projector onto the bit state $0/1$ on site $x$. For example,
\begin{eqnarray}
    \pZ_{x} &=& \mathbb{I}+2\pi^{1}_{x-1}\pi^{1}_{x }\pi^{1}_{x+1} - 2\pi_{x}^{1}\pi_{x+1}^{1}  \nonumber \\
    &&-2 \pi^{1}_{x-1} \pi^{1}_{x} - 2\pi^{0}_{x-1}\pi^{1}_{x }\pi^{0}_{x+1}.
    \label{eqn:z_qp_basis}
\end{eqnarray}
By expanding a region of $O(1)$ sites around $x$ in the computational basis, one can write any operator-state as a sum over $O(1)$ terms, each a with particular pattern of left- and right-moving quasiparticles near point $x$. Since this superposition only involves $O(1)$ terms, we can specialize to one of these terms without changing the asymptotic scaling of the bond dimension. In Appendix \ref{app:Analytic_ancillaries} we work out the details explicitly for the first nontrivial term above; however, the mechanism is generic to local, diagonal operators. To summarize, the operator-state can be written as a sum of $O(1)$ terms, each of which is an equal-weight superposition of computational basis states far from $x$ but projects onto some number of adjacent left and right moving quasiparticles initially centered at $x$, which we refer to as ``marked'' quasiparticles. In what follows we will consider states that have this structure---i.e., a projector onto a particular configuration of ``marked'' quasiparticles on $O(1)$ sites near the operator insertion, and $|\ldots \mathbb{I}\,\mathbb{I}\,\mathbb{I}\ldots \rangle$ far from the insertion.

\emph{Dynamics}.---At short times, it is known~\cite{Klobas_2019,Prosen_Compendium,Alba_2019,Alba_2021} that local operators can be evolved with a bond dimension growing as $ \chi \sim t^{\nu }$ with $\nu = 2$. However, these results only hold at early times relative to system size: they assume that each pair of quasiparticles has collided only once or not at all. This assumption breaks down in each sector at the ``self-scattering time,'' when quasiparticles that started out near each other collide for the second time. As we will see explicitly, the maximum bond dimension across any bipartition of the operator state \emph{projected into a sector} is periodic in the self-scattering time due to the recurrence, so it suffices to consider operator evolution up to that time. However, the self-scattering time is different for each sector, so we have to evolve the operator state separately in each sector. 

\begin{figure}[h]
    \centering
    \includegraphics[width=0.75\linewidth]{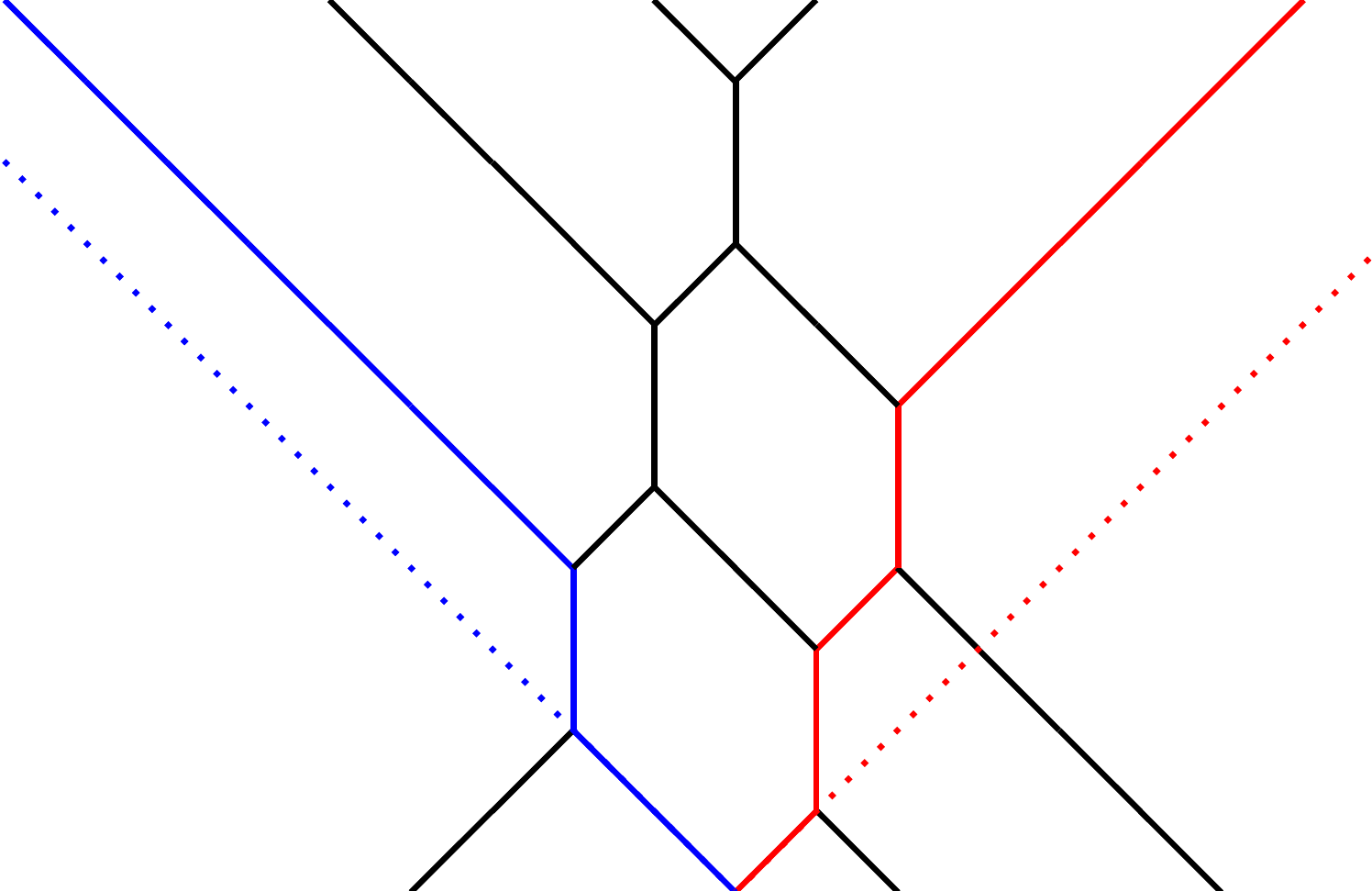}
    \caption{Schematic showing left (blue) and right (red) marked quasiparticles scattering. Dashed lines represent the ``free'' marked quasiparticle worldlines; distances between the solid and dashed lines indicate the accrued time delay. Between the marked quasiparticles, there are two left movers and one right mover, corresponding to a marked left mover with one delay and a marked right mover with two delays.}
    \label{fig:scattering}
\end{figure}

We now discuss what this evolution looks like, in terms of a single term of Eqn.~\ref{eqn:z_qp_basis} that represents a set of marked quasiparticles (for simplicity, one left and one right mover, which capture the key features of the evolution). Crucially, quasiparticles of the same chirality never cross due to the asymptotic spacings being conserved. Thus, the number of time delays experienced by the left/right marked quasiparticle can be counted as the number of right/left movers between the marked quasiparticles. A schematic of this principle is shown in Fig. \ref{fig:scattering}.
Consequently, at a given time $t$, the locations of the marked quasiparticles are fixed by the quasiparticle occupation between them up to an $O\l 1\r$ number of cases (depending on whether they are currently scattering).

We call the region bounded by the left and right marked quasiparticles and containing the initial quasiparticle location $R_{\rm I}$ and its complement $R_{\rm II}$. At time $t$, the marked quasiparticles can encounter $O\l t \r$ quasiparticles each and therefore there are at most $O\l t^{2} \r$ fillings, $\vec{n}$, on $R_{\rm I}$ which correspond with $x_{L/R}$, the locations of the marked left and right mover. Our time-evolved operator state is thus the sum of at most $O \l t^2 \r$ terms: projectors onto the marked quasiparticles at locations $x_{L} = \l x-t + 2n_{R}\r_{L}$ and $x_{R} = \l x + t - 2n_{L}\r_{L}$ with a projector onto $\vec{n}$ quasiparticles between them. Each term has bond dimension of at most $O\l t^2 \r$ (see Appendix~\ref{app:number_projectors}).

\begin{figure}[h]
    \centering
    \includegraphics[width=0.75\linewidth]{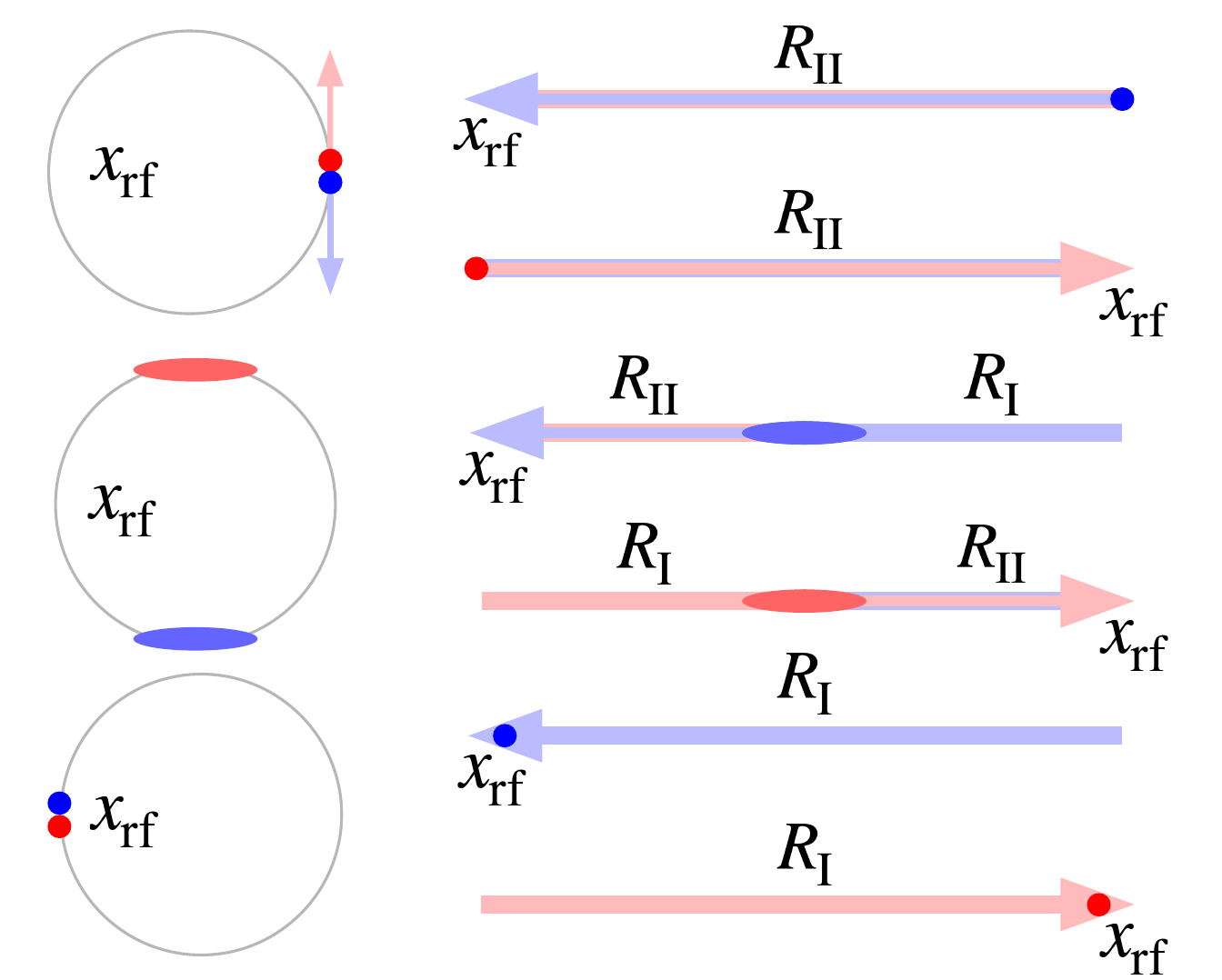}
    \caption{Marked quasiparticle pair refocusing in a fixed-filling sector with $x_{\rm rf}\l \vec{n}\r$ antipodal to $x$. (Top) Initial state of marked quasiparticles at $t=0$. Red and blue striped regions indicate quasiparticles which have yet to scatter with the marked quasiparticles (region $R_{\rm II}$). (Middle) Marked quasiparticle distributions after broadening. All R/L quasiparticles in region $R_{\rm I}$ have scattered with the L/R marked quasiparticle exactly once. All R/L quasiparticles in region $R_{\rm II}$ have not yet scattered with the marked quasiparticles. (Bottom) Quasiparticles arriving at refocusing point with sharp distributions. $R_{\rm II}$ is empty and $R_{\rm I }$ contains $\vec{N}$ quasiparticles.}
    \label{fig:refocusing}
\end{figure}

Now, we consider projecting this time evolved operator state into a given number sector with filling $\vec{N}$ term by term, where each term has a fixed filling $\vec{n}$ on $R_{\rm I}$. The number projector will act trivially on $R_{\rm I}$ (this region is already at fixed filling, $\vec{n}$) and fix the filling on $R_{\rm II}$  to $\vec{N} - \vec{n} - \lb 1,1 \rb$ (where the additional subtraction is for the marked quasiparticles). Thus, the action of the projector is simply to add a number projector onto region $R_{\rm II}$. The positional distribution of the marked quasiparticles will broaden as the number of time delays experienced by them fluctuates. This is illustrated in Fig.~\ref{fig:refocusing}.

Since the total number of time delays in the system is fixed at $\vec{N}$, as the marked quasiparticles traverse the system and as $R_{\rm II}$ becomes smaller the number of possible positions will begin to diminish until the marked quasiparticles collide. We note that the marked quasiparticles will never scatter twice with the same quasiparticle until they have scattered once with every quasiparticle of the opposite chirality. Because the total number of time delays is fixed, the marked quasiparticles will then collide at fixed time and place. Therefore the operator projected into a sector of filling $\vec{N}$ will refocus at position and time
\begin{eqnarray}
    x_{\rm rf} \l \vec{n}\r &=& \l\frac{L}{2} + N_{R}-N_{L} +x  \r_{L} \nonumber \\
    t_{\rm rf}\l\vec{n}\r &=& \frac{L}{2} +N_{L} + N_{R}
    \label{eqn:refocusing_pt}.
\end{eqnarray}
The operator can then be evolved again from its new position within the number sector $\vec{N}$ to arbitrarily late times. Since $t_{\rm rf}$ is $O\l L \r$, we can substitute in $O\l L\r $ for $O\l t\r$ in the bond dimension bounds above.

\emph{Bond dimension counting}.---For each term we get a bond dimension of order $O\l L^{4}\r $ from the two number projectors on $R_{\rm I }$ and $R_{\rm II }$ and have $O\l L^{2}\r $ terms (choices of $x_{L}$ and $x_{R}$). Therefore, the operator state projected into a global number sector can be evolved indefinitely with a bond dimension of $O\l L^{6}\r$. Putting together these results with the cardinality of global number sectors, $O\l L^{2}\r$, we can write our time evolved operator state as a sum across $O\l L^{8}\r $ factorized (between any $A$ and $\bar{A}$) terms. This leads to the bound
\begin{equation}
    S_{\rm vN} \leq S_{\l 0 \r} \lesssim 8 \, {\rm Log}_{2}\lb L \rb
\end{equation}
for all times, where we have exploited the fact that the Hartley entropy (which is the logarithm of the bond dimension) upper bounds the von Neumann entropy. This bound is expected to be very loose in large part because number fluctuations are central limiting (rather than flatly distributed).

\subsection{Numerical evidence}

\label{sect:Rule54_LT}
\begin{figure}[h]
    \centering
    \includegraphics[width =0.75 \linewidth]{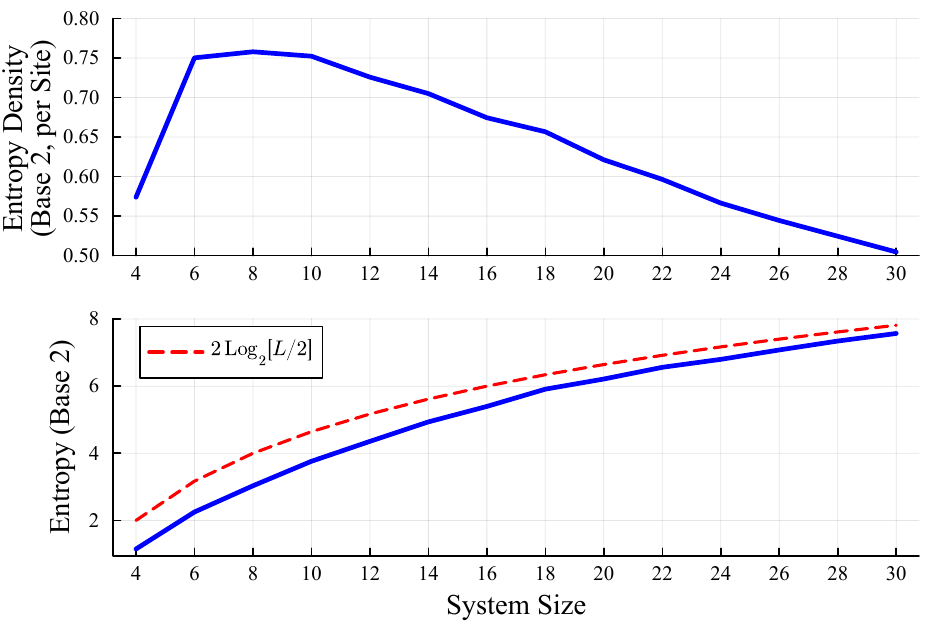}
    \caption{Late-time averaged LOE of single-site $\pZ$ operator with half-space bipartition. (Top) LOE entropy density in base 2 units (bits) per site. (Bottom) Operator entropy as a function of system size. Numerical results are compared to $2 \, {\rm Log}_{2}\lb L /2\rb$.}
    \label{fig:diagonal_log_law}
\end{figure}
In accordance with the theoretical predictions of Section \ref{sect:diag_analytics}, numerical evidence indicates that diagonal operators saturate to logarithmic scaling rules at long times per Fig. \ref{fig:diagonal_log_law}. The time average shown in this figure is drawn from a sampling of $11^{10}+ 37n $ \footnote{This expression scans as a perfect line of iambic hexameter broken in the middle-- otherwise known as an alexandrine.} Floquet pumps ($\delta t  = 2$) with $n \in \mathbb{N} \,  \cap \, \lb 0 , 99 \rb$. The lateness of the initial time is unnecessary (though shorter time samples are in full agreement), but the fact that the timestep is prime is of significant import. The density of states for system size $L$ in Rule 54 has spikes at  multiples of $2\pi /L$, so times which are sufficiently large multiples of $L$ will experience strong recurrences.

Though the results of Fig. \ref{fig:diagonal_log_law} are shown in bits or bits per site, we note that the most saturated system size still has less than half its possible maximum entropy since it is diagonal. The deviation of the maximum entropy density from or $1$ in bits per site is consistent with a Page correction \cite{Page_1993}. We note the strong agreement with $2\log{L/2}$, as shown in Fig.~\ref{fig:diagonal_log_law}. The operator $\pZ_{1}\pZ_{2}$ produces qualitatively identical results, but adding a trace to the initial operator starkly changes the saturation, as expected \cite{Alba_2024}.

Probing the saturation of off-diagonal operators is considerably more difficult numerically. R\'enyi entropies are not reliable indicators of the von Neumann entropy, due to substantial overlap with conserved quantities (see also \cite{Bertini_2022}). We can gain additional insights by considering smaller bipartitions, but we are limited to system sizes of $L\leq 16$ for the half-space bipartition which registers deviations from volume law at the smallest system sizes. For $L\div 3$ and $L\div 4$, we must account for strong parity effects.

\begin{figure}[h]
    \centering
    \includegraphics[width =0.75 \linewidth]{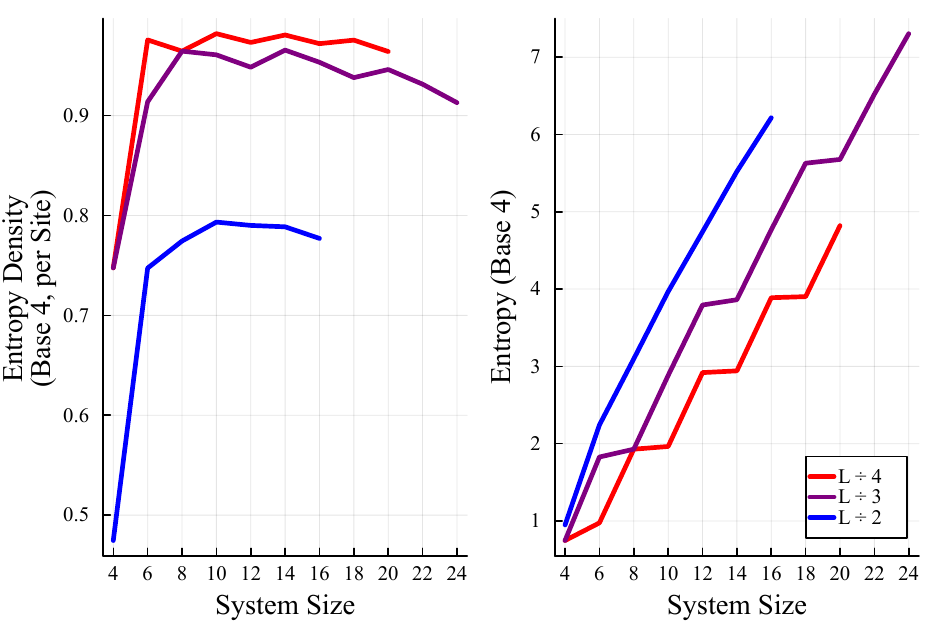}
    \caption{Late-time averaged half, third, and quarter space partitioned LOE of a single-site, off-diagonal $\pX$ operator. (Left) Entropy densities in base $4$ units per site for half, third, and quarter space bipartitions. (Right) Entropies in base 4 units. Appendix \ref{app:memory} explains why we attain larger system sizes for the $L\div 3$ partition than the $ L\div 4$ partition.}
    \label{fig:off_diagonal}
\end{figure}

We select traceless, single-site initial operator $\pX$. Half/third-space bipartitions for the two-site operator $\pX_{L} \pX_{1}$ were also examined, showing very similar results (much like the case of diagonal operators). Long time averages are shown in Fig. \ref{fig:off_diagonal}.
It is challenging to immediately interpret this data, but it is evident that there is a downturn in the half-space bipartite entropy density for system sizes between $L= 10$ and $L=14$. Additionally, there is an evident downturn in the less sensitive third-space bipartition entropy density for system sizes greater than $L=18$. 

Fig.~\ref{fig:off_diagonal} suggests that off-diagonal operators in Rule 54 will also saturate to sub-volume laws, like diagonal operators. Indeed, off-diagonal operators refocus like diagonal operators (see Appendix~\ref{sect:od_analytics}), though with a timescale $O\l L^2 \r$ that makes analogous logarithmic bounds challenging to establish.  Again, for the  maximum half-space entropy density at $L=10$, the deviation from $1$ is consistent with a Page correction.

\section{LOE for General Integrable Models}
\label{sect:gen_loe}

Rule 54 is simpler than generic integrable models in a few different ways: (i)~there are only two types of quasiparticles, because of the lack of dispersion, (ii)~for the same reason, it is possible to write down diagonal operators that remain diagonal in the computational basis at all times, and (iii)~the expansion of a local operator in the quasiparticle basis is relatively simple. General integrable models, like the Heisenberg model, have none of these features: instead, they have dispersive quasiparticles, so there are formally infinitely many different quasiparticle types, parameterized by a continuous ``rapidity'' label. (There might be additional discrete ``string'' indices but these will not matter for our analysis.) Moreover, when two quasiparticles collide, the scattering phase shift generally depends on both rapidities. Therefore, the ``refocusing'' phenomenon discussed in Rule 54 does not take place in general. 

In this section we will argue informally that these differences lead to a parametrically larger saturated value of operator entanglement in generic interacting integrable models. Our discussion will have three parts. First, we will introduce a toy model (which we call the $K$-flavor model) that generalizes Rule 54 to the case of many inequivalent flavors, and use this model to argue for volume-law scaling of the LOE in general interacting integrable systems. Second, we will provide numerical support for volume-law scaling of the late-time LOE in the Heisenberg model. Third, we will discuss why free fermions, despite having nontrivial dispersion, evade this argument for volume-law entanglement. (It is known~\cite{Dubail_2017,Rath_2023, Murciano_2023} that local operators have either constant or logarithmic entanglement in these systems.)

\subsection{$K$-Flavor model}
\label{sect:heisenberg_analytics}

A generic interacting integrable model like the Heisenberg spin chain has two key features---(i)~the existence of dispersive quasiparticles and (ii)~the rapidity-dependence of the scattering phase shifts---that distinguish it from Rule 54. These features also make such models intractable. In this section, instead, we study a toy model that captures feature~(ii), by generalizing Rule 54 to a model with $K$ inequivalent flavors of quasiparticles. Eventually, we will take $K \sim N$. Each of the $K$ flavors is still nondispersive, and the $K$ scattering phase shifts (that a left-mover of one flavor has with all right-movers) are taken to be incommensurate real numbers. Thus, as phrased, the $K$-flavor model does not have a lattice realization; it can be regarded as a nondispersive hard-rod gas in the continuum. However, one can still define and analyze diagonal operators (which map to states) exactly as in Rule 54.

In the $K$-flavor model, a marked quasiparticle dephases over time between states with distinct vectors $\vec{N} = (N_1 \ldots N_K), \sum_{\alpha=1}^K N_\alpha \leq L$. Therefore, in the late-time limit, the operator density matrix is approximately diagonal in $\vec{N}$, and its entropy scales at least with the Shannon entropy of the probability distribution of the infinite temperature state over $\vec{N}$ sectors.

Then, the (categorical) sector probability distribution is
\begin{eqnarray}
    p\l \lcb N_{\alpha} \rcb\r &=& \frac{1}{\mc{Z}} \ \Theta \l L -  \sum_{\alpha} N_{\alpha }\r \prod_{\alpha = 1 }^{K} \binom{L}{N_{\alpha}} \nonumber \\
    &&\mc{Z} = \sum_{N = 0}^{L}\binom{KL}{N}.
\end{eqnarray}

To understand the entropy of this distribution, we can leverage a saddle point at $O\l 1 \r$ occupation per flavor. A potentially competing contribution from $O\l L \r$ filling in $O\l 1 \r$ flavors (due to the non-trivial multiplicities) can be ruled out. Around this saddle point, the distribution can be approximated by the (equal-probability) multinomial distribution: let $\sum_{\alpha}N_{\alpha} = N$, then
\begin{equation}
    p\l N;  \lcb N_{\alpha}\rcb \r = \frac{N! K^{-N}}{\prod_{\alpha}N_{\alpha}!}.
\end{equation}
A general closed form expression for the entropy of this distribution is not known; however, in the case of $N\to \infty $ and $K\to \infty$, with $\rho \equiv N/K $ held constant and $O\l 1 \r$, the filling in each different flavor can be made independent to a good approximation and the entropy scales with $L$. An intuitive way to see this is that each flavor acts as an independent degree of freedom, implying that entropy scales as $K$ multiplied by an order one constant related to $\rho$. This captures one of the two terms, and the other is proportional to the filling, $N$. The calculation is equivalent to the free energy of a lattice Bose-Einstein gas with occupation $N$ and $K$ sites, for which we provide an exact formula in Appendix \ref{app:Saddle_point}, in addition to other relevant calculations.
\color{black}

\subsection{Heisenberg model}

The dephasing argument we outlined for the $K$-flavor model works \emph{a fortiori} for the Heisenberg model, which has the same structure of scattering phase shifts, as well as dispersing quasiparticles (which add a further dephasing mechanism). On these grounds, therefore, we would expect the Heisenberg model to exhibit a volume-law LOE at late times. Note that our analysis of the $K$-flavor model does not depend in any crucial way on the fact that quasiparticles can be created locally. Provided that the excitation creates or ``tags'' quasiparticles, the motion of these quasiparticles through an infinite-temperature state is sufficient to cause the dephasing that we need to argue for a volume law. 

\begin{figure}[h]
    \centering
    \includegraphics[width = 0.75\linewidth]{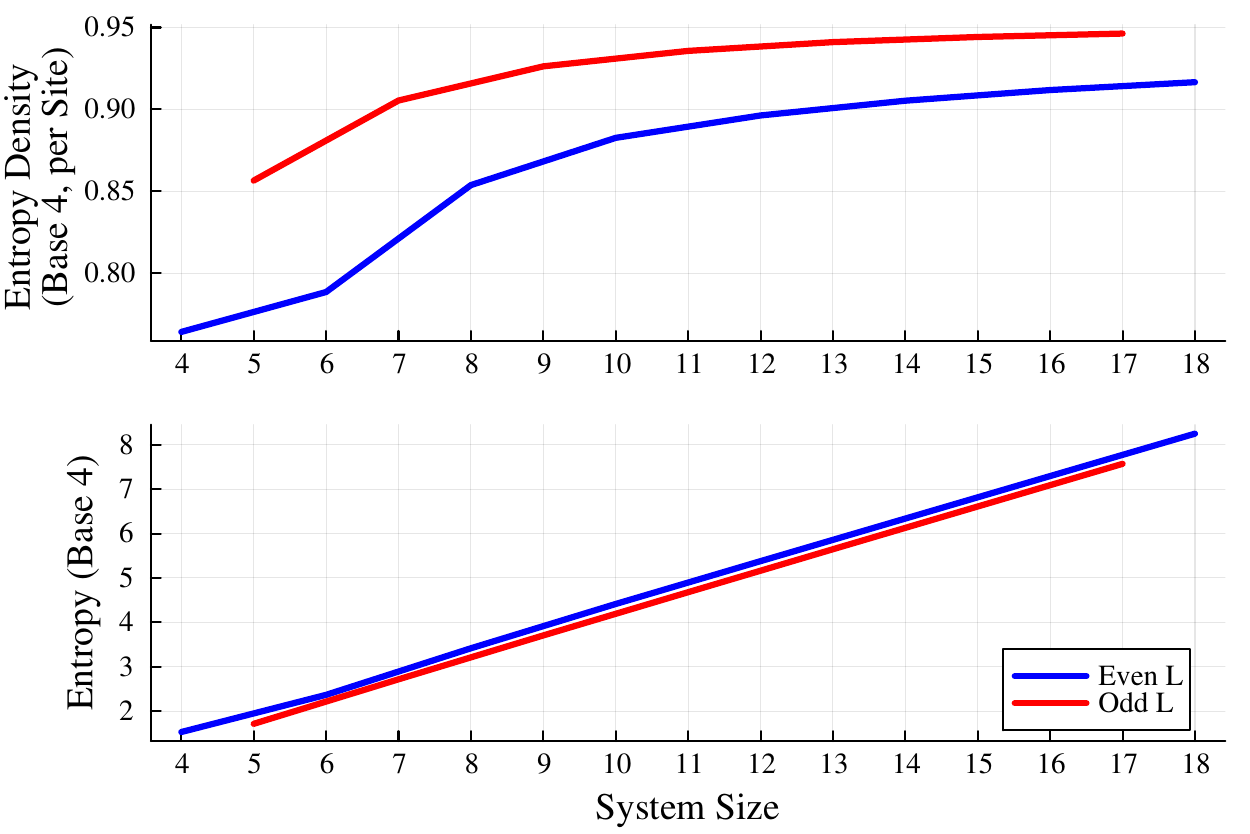}
    \caption{Time averaged LOE of $U(1)$ projected $\pZ_{1} \pZ_{2}$ at late time with half-space bipartition. (Top) Operator entropy density (base $4$ units, per site). (Bottom) Operator entropy (base $4$). Data was sampled from $t=500$ to $t =600$ in intervals of $5$ time units.}
    \label{fig:Heisenberg_Vol_Law}
\end{figure}

We now numerically examine the saturation behaviour of the Heisenberg model. For system sizes beyond approximately $L=14$ it is challenging to both time evolve to sufficiently late times and extract entropy for sufficient time samples, since exact diagonalization is needed for both tasks. To attain larger system sizes in the long time limit we make use of a $U(1)$ projection and symmetry resolution technique, detailed  in Appendix \ref{app:symmetry_resolution}; similar symmetry resolutions have recently appeared in the literature \cite{Rath_2023,Murciano_2023}. Results are shown for systems sizes of $L=4$ to $L=18$ in Fig. \ref{fig:Heisenberg_Vol_Law}, using initial operator $\pZ_{1}\pZ_{2}$ with a half-space bipartition. The operator is projected into the half-filling $U(1)$ charge sector; this appears to have little impact upon the entropy density at late times for the system sizes which could be verified by full Hilbert space exact diagonalization.
The numerical results from this symmetry projection scheme show a sharply-defined volume-law, with non-decreasing entropy density in system size. The distinction between even and odd system sizes is significant as the dimension of the $U(1)$ charge sector which dominates the entropy has a pronounced even-odd effect (that is, the even-odd effect is not merely an artifact of the symmetry resolution). 

To verify the validity of numerics in the $U(1)$ symmetry resolution scheme, the differences with the full Hilbert space results for system sizes up to and including $L=13$ is shown in Fig. \ref{fig:u1_comparison} of Appendix \ref{app:u1_comparison}. Differences are significant at early times for small $L$, but diminish in $L$ (though the time to which they remain increases in $L$-- averages are taken at sufficiently late times that this is not a concern for data shown in Fig. \ref{fig:Heisenberg_Vol_Law}). This model exhibits considerably smaller, and exponentially diminishing in $L$, fluctuations in operator entropy at late times.

\subsection{What about free fermions?}

Models such as the XX and transverse-field Ising spin chains provide an interesting intermediate case: they have dispersive quasiparticles (and therefore exponentially many ``sectors'') but are simple enough that direct calculations of the LOE are possible \cite{Dubail_2017,Rath_2023,Murciano_2023}. These calculations (which use very different methods than the semiclassical reasoning here \cite{Peschel_03,Peschel_09}) show that there are two types of operators: parity-even operators, which are bilinear in the fermionic quasiparticles, and have $O(1)$ LOE; and parity-odd operators, which involve a Jordan-Wigner string in the fermionic representation, and have $O(\log t)$ growth of LOE. The basic distinction between these free models and interacting ones is that the dephasing arguments on which we are relying do not apply: the propagator for a fermionic quasiparticle is state-independent, so coherences of the operator between states with distinct quasiparticle content cannot be neglected. 

\section{Discussion}
\label{sect:conclusion}

In this work we investigated the saturation of LOE in the Rule 54 and quantum Heisenberg chains. For Rule 54 we derived late time bounds for the LOE of diagonal operators and performed a robust numerical analysis for off-diagonal operators which indicated that LOE will saturate logarithmically. For the quantum Heisenberg chain, we conjectured volume-law saturation, and supported this conjecture with numerical evidence. 

\subsection{R\'enyi entropies}

So far, we have focused on the Von Neumann entropy. It is straightforward to see that operator R\'enyi entropies with $\alpha > 1$ are never volume-law in the models we are considering; indeed, they are generically not volume-law for operators in any model with conservation laws. The argument for this is simple: if one expands the operator at a generic very late time in an operator basis that includes the conserved charge, its overlap onto the conserved charge will generically be polynomially small in $1/L$. Thus the largest Schmidt coefficient of the operator-state will scale polynomially in $1/L$, immediately implying that the min-entropy (and therefore all R\'enyi entropies with $\alpha > 1$) are at most logarithmic in $L$. 

\subsection{Time dynamics}

Numerical results on saturation timescales and intermediate times are provided in Appendix~\ref{app:early_time}. For Rule 54, our numerical results suggest that the entropy saturates on the self-scattering timescale: namely, $t \sim L$ for diagonal operators and $t \sim L^2$ for off-diagonal operators. For the Heisenberg model, our numerical results on the saturation timescale are inconclusive. 

The dephasing picture, however, suggests a natural conjecture for the temporal growth of entanglement. After a time $t$, a marked quasiparticle has spread out through collisions over a distance $t^{1/2}$. However, its quantum mechanical broadening over this timescale is $t^{1/3}$, so its position does not distinguish between collision histories that gave the same phase shift to within this resolution. Thus the quasiparticle position only carries $\sim (1/6) \log t$ bits of information about its collision history, consistent with the observed logarithmic growth of entanglement. The dynamics of the crossover between the early and late-time regimes of entanglement growth is an interesting topic for future work. 

\subsection{Other directions}

There are a few avenues by which our conjecture for volume laws in generic interacting integrable models could be made more rigorous or contradicted. A particularly intriguing avenue is via the so-called no-resonance theorems \cite{Linden_2009, Kaneko_2020, huang_2021, Mark_2022, Riddell_2023, Riddell_2024}, which may provide a rigorous motivation for dephasing, which would be sufficient to demonstrate volume law entanglement. The differences between the interacting single-body dispersive and non-dispersive cases are well motivated in this picture since in interacting, single-body dispersive systems if time delays depend non-trivially on the rapidity of scattering quasiparticles, the spectrum is far likelier to be incommensurate, since the time delay function must be fine-tuned. However, this point raises the question of whether single-body dispersion and uniform interactions are sufficient to achieve volume law LOE, which we do not resolve.

While we are not aware of any numerical techniques that can handle larger system sizes at late times than those given for the Rule 54 chain in this study, a large scale numerical study of the model proposed in \cite{Friedman_2019} (a dispersive generalization of the Rule 54 chain) could potentially be illuminating. The central difficulty of such a study is that this model has a six-site update rule and larger system sizes would be needed to study the scaling behaviour of LOE with system size. If the saturation of this model were understood, it might be sufficient to resolve the above ambiguity (whether uniform time delays in addition to single body dispersion are sufficient). Additional studies of LOE on finite chains with integrability preserving boundary conditions might also prove insightful.
\section*{Acknowledgements}
    We would like to highlight the assistance of David Huse, with whom we had many helpful discussions throughout the duration of this work. We would also like to thank Bruno Bertini, Ross Dempsey, Katja Klobas, Pavel Kos, Samuel J Li, Daniel Mark, and Marko \v{Z}nidari\v{c} for insightful comments and/or stimulating discussions.


\paragraph{Funding information}
J.A.J. was partially supported by the National Science Foundation Graduate Research Fellowship Program under Grant No. DGE-2039656. S.G. was partially supported by NSF QuSEC-TAQS OSI 2326767. Any opinions, findings, and conclusions or recommendations expressed in this material are those of the authors and do not necessarily reflect the views of the National Science Foundation. The simulations presented in this article were performed on computational resources managed and supported by Princeton Research Computing, a consortium of groups including the Princeton Institute for Computational Science and Engineering (PICSciE) and the Office of Information Technology's High Performance Computing Center and Visualization Laboratory at Princeton University. We gratefully acknowledge their hard work, maintaining and improving these computational resources.

\begin{appendix}
\numberwithin{equation}{section}

\section{Operator Support and Operator States}
\label{app:notation_and_support}

\subsection{Why OTOCs Measure Operator Support}
\label{app:otocs}
A standard tool to probe operator support (region on which the operator acts not as the identity) is an Out of Time-Order Correlator/Commutator (OTOC). The  OTOC is defined as
\begin{equation}
    \mc{C} \l x \r = \frac{1}{8} \sum_{k=1}^{3} \lv \lb   O, \sigma^{k}_{x}\rb\rv^2 .
\end{equation}
where we assume this operator to be Hilbert Schmidt normalized ($\lv O\rv^{2} = \tr{O^{\dagger}O}=1$).

Let us denote the subset of normalized Pauli strings that act as a non-identity on site $x$ as $\hat{S}_{x}$. We can separate the Pauli string decomposition into two parts for a given site $x$
\begin{eqnarray}
    O = \sum_{\hat{S} \in \hat{S}_{x}} a_{\hat{S}} \hat{S} \, +\,  {\rm Everything \ Else}. 
\end{eqnarray}
The OTOC then measures the fraction of $O$ that acts non-trivially as
\begin{equation}
    C\l x, t \r  = \lv\sum_{\hat{S} \in \hat{S}_{x}} a_{\hat{S}} \hat{S}\rv^{2} = \sum_{\hat{S} \in \hat{S}_{x}} \lv a_{\hat{S}}\rv^{2}.
\end{equation}
This quantity can be extracted by taking the commutator with Pauli matrices since the local Pauli only anticommutes with non-identity entries in the string, $\hat{S}$. 

\subsection{Operator Entropy}
\label{sect:operator_entropy}

Equivalent to the definition in the main text, we can also define operator entanglement in the pure operator formalism
\begin{equation}
    \hat{O} = \sum_{k}\sqrt{F_{k}} \ \hat{O}^{A}_{k} \otimes \hat{O}^{B}_{k}
    = 
    \sum_{k}\sqrt{F_{k}} \ \frac{O^{A}_{k} \otimes O^{B}_{k}}{\lv O^{A}_{k}\rv \lv O^{B}_{k}\rv}
    \label{eqn:op_schmidt}
\end{equation}
where $\hat{O}$ is Hilbert-Schmidt normalized. By exact analogy with the state case, we then define the operator entropy on region $\rm A$ in base-4 units as
\begin{equation}
    \svn{O;\, {\rm A}} = -\sum_{k} F_{k} {\rm Log}_{4}\lb F_{k}\rb .
\end{equation}

A slightly counterintuitive feature of operator entropy is that the product of two operators do not have additive entropy. Consider two operators $\hat{O}$ and $\hat{Q}$, for the vast majority of choices $\lv \hat{O} \hat{Q}\rv^{2} \neq 1 $. Then, we find that
\begin{eqnarray}
    \frac{OQ}{\lv O Q\rv}  = \sum_{j, k }&&\sqrt{\frac{F_{k}D_{j} \lv  \hat{O}^{A}_{k}\hat{Q}^{A}_{j}\rv^{2} \lv  \hat{O}^{B}_{k}\hat{Q}^{B}_{j}\rv^{2} }{\lv OQ \rv^{2} }}  \nonumber \\
    && \times \frac{\hat{O}^{A}_{k}\hat{Q}^{A}_{j}}{\lv \hat{O}^{A}_{k}\hat{Q}^{A}_{j} \rv}\otimes \frac{\hat{O}^{B}_{k} \hat{Q}^{B}_{j}}{\lv \hat{O}^{B}_{k} \hat{Q}^{B}_{j}\rv}
\end{eqnarray}
where the Schmidt bases of operators $O$ and $Q$ are labeled by $k$ and $j$ respectively. The Schmidt basis of $OQ$ has cardinality upper bounded by the product of the cardinalities of the Schmidt bases of $O$ and $Q$, and therefore the Hartley entropy of the product $OQ$ is upper bounded by the sum of the Hartley entropy of $O$ and $Q$. However, other entropy metrics do not have straightforward inequalities. The essential difficulty comes from the factor of $\lv O Q\rv$ which is needed to normalize the product.

\section{Ancillaries for Analytic Results}
\label{app:Analytic_ancillaries}

\subsection{Time Evolution}
\label{app:diag_anc}
Here we present a proof of the bounds described in Section \ref{sect:diag_analytics}, examining in detail the structure of diagonal operator support in Rule 54. The derivation is mostly a minor deviation from previous work, per \cite{Alba_2019, Alba_2021}. We will work in the operator formalism for notational convenience.

\begin{figure}[h]
    \centering
    \includegraphics[width=0.75 \linewidth]{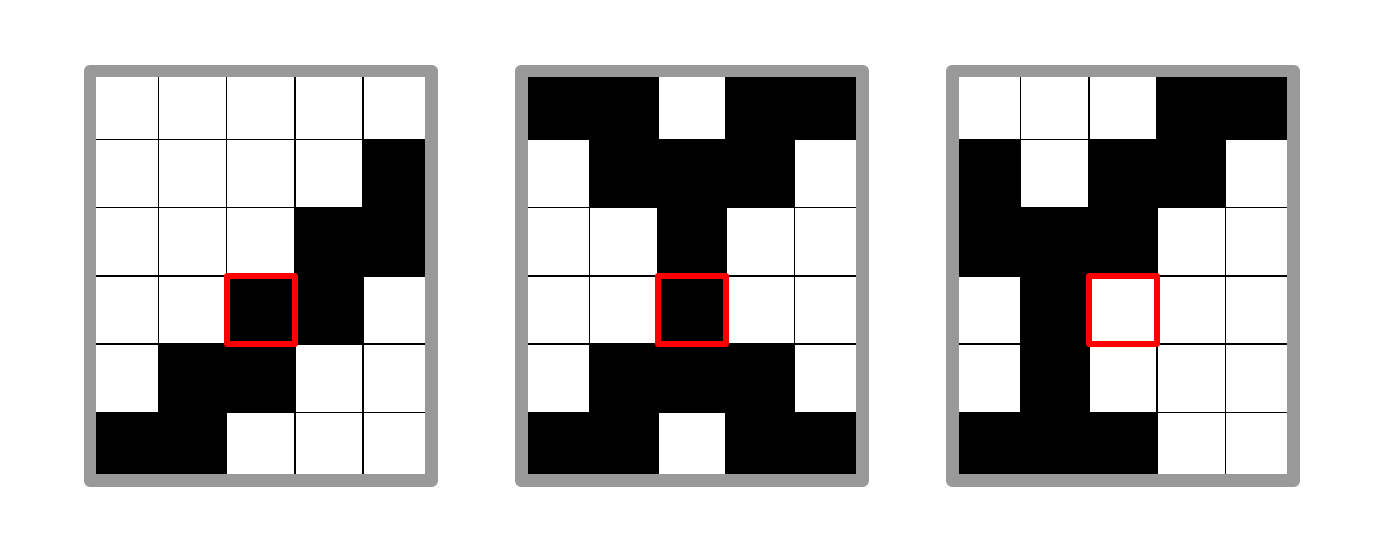}
    \caption{Three cases for a right mover at $x_{R}$ (marked in red). (Left) free, $\mc{I}^{R}_{x}$ (Middle) first delay, $\mc{S}_{x}$ (Right) second delay, $\mc{S}_{x-1}$. The first and second delayed quasiparticles will always have the same (distinct) relative spatial shifts compared to their free counterpart. After the second delay, the number projector on $R_{\rm I}$ will pick up the left mover and take $x_{R}\to x_{R}-2$.}
    \label{fig:scattering_cases}
\end{figure}

We start with $\pi^{1}_{x-1}\pi^{1}_{x}\pi^{1}_{x+1}$ as our initial operator at $t=0$. We assume the even/odd pump order of Eqn.~\ref{eqn:U_R54} and $x$ to be odd, though the even case is simply related by symmetry (see Eqn.~\ref{eqn:time_reversal}). We define a set of projectors, which we call ``marked projector'' operators. These are
\begin{equation} 
    \mc{I}^{R/L}_{x} = \pi^{1}_{x}\pi^{1}_{x\pm 1 } \quad \mc{S}_{x} = \pi^{0}_{x-1} \pi^{1}_{x} \pi^{0}_{x+1}
\end{equation}
so as to project onto an isolated and scattering quasiparticle respectively.  It is convenient to adopt the notation that an R/L quasiparticle ``lives'' on its left/rightmost $1$. The time delay consists of two timesteps which are disambiguated by the lattice parity (even/oddness) of the site upon which the scattering quasiparticles sit with respect to the even/oddness of the timestep; at a given time $\mc{S}_{x}$ and $\mc{S}_{x+1}$ imply the quasiparticle pair are at different stages of a scattering event, per Fig.~\ref{fig:scattering_cases}. A marked quasiparticle's location is fixed up to three cases by its scattering history --- for $\pi^{1}_{x-1}\pi^{1}_{x}\pi^{1}_{x+1}$ the number of quasiparticles of the opposite chirality between them (on $R_{\rm I }$ per the main text).

For isolated quasiparticles, we can define the time delayed shift and position as
\begin{eqnarray}
    \delta x_{L/R}\l t,n_{R/L} \r &=&  t -2 n_{R/L}  \nonumber \\
    x_{L/R}\l t , n_{R/L}\r &=&\l x \mp \delta x_{L/R} \l t,  n_{R/L}\r \r_{L}.
    \label{eqn:position_of_number}
\end{eqnarray}
Per the main text, we consider projecting into an overall number sector with $\vec{N}$ quasiparticles. Above we defined the quasiparticle occupation projector between $\l x, y\r$, $\pi_{\vec{n}}^{\l x, y\r}$, as the operator counting quasiparticles between a quasiparticle at $x$ and a quasiparticle at $y$ not including the endpoints. This operator is defined explicitly in Eqn.~\ref{eqn:number_counting}. It is not defined in the absence of quasiparticles which ``live'' at its endpoints $x$ and $y$, or more structure fixing the boundary bits. We note that $\l y, x\r$ denotes the open complement of $\l x, y\r$.

The operator can be evolved by projecting onto all possible locations of the marked quasiparticles originating from the pattern $1_{x-1}1_{x}1_{x+1}$ with all possible fillings consistent with that location between them (on $R_{\rm I }$) and all possible fillings consistent with the total filling constraint from the global number projector beyond them (on $R_{\rm II }$). There are three cases at any given filling for each marked quasiparticle (shown in Fig.~\ref{fig:scattering_cases}), which leads to nine terms in the sum. The result is
\begin{eqnarray}
    & &\pi_{\vec{N} }U^{\dagger}\l t \r\pi^{1}_{x-1}\pi^{1}_{x }\pi^{1}_{x+1}U\l t \r  \pi_{\vec{N} } =\pi_{\vec{N} }U^{\dagger}\l t \r\pi^{1}_{x-1}\pi^{1}_{x }\pi^{1}_{x+1}U\l t \r  \nonumber \\
    =\sum_{\vec{n}} \bigg{(}&& \mc{I}_{x_{L}\l t,\vec{n} \r}^{L}    \pi^{\l x_{L}\l t , \vec{n}\r, x_{R}\l t, \vec{n}\r\r}_{\vec{n}}     \mc{I}_{x_{R}\l t,\vec{n} \r}^{R} \pi^{\l x_{R}\l t, \vec{n}\r + 1 ,  x_{L}\l t,  \vec{n}\r -1 \r}_{\vec{N} - \vec{n} - \lb 1,1\rb} \nonumber \\
    &&+ \mc{I}_{x_{L}\l t,\vec{n} \r}^{L}    
    \pi^{\l x_{L}\l t , \vec{n}\r, x_{R}\l t, \vec{n}\r\r}_{\vec{n}}     
    \mc{S}_{x_{R}\l t,\vec{n} \r}^{R} 
    \pi^{\l  x_{R}\l t, \vec{n} \r,  x_{L}\l t,  \vec{n}\r -1  \r}_{\vec{N} - \vec{n} - \lb 2,1\rb} \nonumber \\
    &&+ \mc{I}_{x_{L}\l t,\vec{n} \r}^{L}    
    \pi^{\l x_{L}\l t , \vec{n}\r, x_{R}\l t, \vec{n}\r -1\r}_{\vec{n} }     
    \mc{S}_{x_{R}\l t,\vec{n} \r - 1 }^{R} 
    \pi^{\l x_{R}\l t , \vec{n} \r -1,  x_{L}\l t , \vec{n}\r -1 \r }_{\vec{N} - \vec{n} - \lb 2,1\rb} \nonumber \\
    &&+ \mc{S}_{x_{L}\l t,\vec{n} \r}^{L}    
    \pi^{\l x_{L}\l t , \vec{n}\r, x_{R}\l t, \vec{n}\r\r}_{\vec{n}}     
    \mc{I}_{x_{R}\l t,\vec{n} \r}^{R} 
    \pi^{\l  x_{R}\l t , \vec{n} \r + 1, x_{L}\l t , \vec{n}\r  \r}_{\vec{N} - \vec{n} - \lb 1,2\rb} \nonumber \\
    &&+ \mc{S}_{x_{L}\l t,\vec{n} \r}^{L}    
    \pi^{\l x_{L}\l t , \vec{n}\r, x_{R}\l t, \vec{n}\r\r}_{\vec{n}}     
    \mc{S}_{x_{R}\l t,\vec{n} \r}^{R} 
    \pi^{\l  x_{R}\l t,  \vec{n} \r,  x_{L}\l t,  \vec{n}\r \r}_{\vec{N} - \vec{n} - \lb 2,2\rb} \nonumber \\
    &&+ \mc{S}_{x_{L}\l t,\vec{n} \r}^{L}    
    \pi^{\l x_{L}\l t , \vec{n}\r, x_{R}\l t, \vec{n}\r -1 \r }_{\vec{n}}     
    \mc{S}_{x_{R}\l t,\vec{n} \r-1 }^{R} 
    \pi^{\l x_{R}\l t, \vec{n}\r-1 ,  x_{L}\l t, \vec{n}\r \r}_{\vec{N} - \vec{n} - \lb 2,2\rb} \nonumber \\
    &&+ \mc{S}_{x_{L}\l t,\vec{n} \r -1 }^{L}    
    \pi^{\l x_{L}\l t , \vec{n}\r -1 , x_{R}\l t, \vec{n}\r\r}_{\vec{n}}     
    \mc{I}_{x_{R}\l t,\vec{n} \r}^{R} 
    \pi^{\l x_{R}\l t, \vec{n} \r +1 ,  x_{L}\l t, \vec{n} \r-1  \r}_{\vec{N} - \vec{n} - \lb 1,2  \rb} \nonumber \\
    &&+ \mc{S}_{x_{L}\l t,\vec{n} \r -1 }^{L}    
    \pi^{\l x_{L}\l t , \vec{n}\r -1 , x_{R}\l t, \vec{n}\r\r}_{\vec{n}}     
    \mc{S}_{x_{R}\l t,\vec{n} \r}^{R} 
    \pi^{\l x_{R}\l t,  \vec{n} \r, x_{L}\l t, \vec{n}  \r-1   \r}_{\vec{N} - \vec{n} - \lb 2,2  \rb} \nonumber \\
    &&+ \mc{S}_{x_{L}\l t,\vec{n} \r -1 }^{L}    
    \pi^{\l x_{L}\l t , \vec{n}\r -1 , x_{R}\l t, \vec{n}\r - 1 \r}_{\vec{n}}     
    \mc{S}_{x_{R}\l t,\vec{n}   \r -1 }^{R} 
    \pi^{\l x_{R}\l t, \vec{n} \r -1, x_{L}\l t, \vec{n}  \r-1    \r}_{\vec{N} - \vec{n} - \lb 2,2  \rb} \bigg{)}
    \label{eqn:diag_ev_1}
\end{eqnarray}
where $x_{L/R}\l\vec{n}\r \equiv x_{L/R}\l t , n_{R/L}\r =\l x \mp \delta x_{L/R} \l t,  n_{R/L}\r \r_{L} $ per Eqn~\ref{eqn:position_of_number}. The entropic cost of this superposition to account for the cases is an $O\l 1\r$ constant \cite{Linden_06,Gour_07,Gour_08}. We define the number projectors carefully in the next section. Since the number projectors have bond dimension of at most $O\l L^2 \r$ each; there are $O\l L^2 \r$ terms for each number sector; and there are $O\l L^2\r$ sectors, the bond dimension of the evolved operator is at most $O\l L^{8}\r$ at leading order per the main text. This bond dimension of the operator as an MPO is exactly equivalent to the bond dimension of the operator as an MPS in the state mapping by definition.

For other diagonal operators we expect the casework to be more complicated, but the procedure fundamentally the same (as we argued in the main text). The initial operator can be broken into the quasiparticle basis locally. Then, each term can be understood as two local formations of left and right movers with fixed asymptotic spacings. Each term can then be time evolved by counting the number of time delays experienced by the whole formation (quasiparticles between the formations) and splitting into cases (exponentially many in the initial size of the operator) which account for the current scattering status of the quasiparticles in the formation. Since the number of cases is only exponential in the support of the initial operator in the quasiparticle basis, the casework can only introduce an $O\l 1 \r$ correction to LOE for initially local operators.

\subsection{Number Projectors, Marginals of Number Projectors, and Bulk Support}
\label{app:number_projectors}
To construct the number projectors of Rule 54, we begin by reviewing (and slightly modifying) the number counting scheme which is given, for example, in the supplementary material of \cite{GopalakrishnanR54-2}. Particularly, we want to define number operators on a region between $x_{L}$ and $x_{R}$, $A = \l x_{L}, x_{R}\r$; it is helpful to define the regional even and odd sublattices as $R_{\rm e/o } = A\cap \mathscr{L}_{\rm even/odd}$. Then the number operators on $A$ can be expressed as  
\begin{equation}
    n^{A}_{R/L}\l t \r = \sum_{x \in \l x_{L}, x_{R}\r} \pi_{x-1}^{0}\pi_{x}^{1}\pi_{x+1}^{0}+ 
    \begin{cases}
        \sum_{x\in R_{\rm o}} \pi_{x}\pi_{x\pm 1 } & t \  {\rm even } \\
        \sum_{x\in R_{\rm e}} \pi_{x} \pi_{x\pm 1 }& t \  {\rm odd }
    \end{cases}
    \label{eqn:number_counting}
\end{equation}
where the even and odd time distinction assumes the pump ordering of Eqn. (\ref{eqn:U_R54}). The projectors will ``hang over the edge'' onto sites $x_{L/R}$. We define the number projectors as 
\begin{equation}
    \pi_{\vec{n}}^{A} = \sum_{\lcb \phi^{k}  \vert \vec{n}\l \phi^{k},t\r = \vec{n}\rcb} \lv \phi^k \ket \bra \phi^k \rv 
\end{equation}
where $\phi^{k}$ are bitstrings on $\lb x_{L}, x_{R}\rb$. Formally, when supported on regions with boundaries, the number projectors are ill-defined at their boundaries; generally, we need to fix the last boundary bits at $x_{L/R}$, to accurately count quasiparticles on a region. This pathology is cured by the marked quasiparticle operators in the previous section (which project out the ``wrong'' boundary indices). We now concern ourselves with how to Schmidt decompose these number projectors.

A priori, it appears challenging to define a low bond-dimension MPO to project onto a given occupation. To do this it is helpful to consider the number projector with fixed boundary bits
\begin{equation}
      \pi_{\vec{n}}^{A} \pi^{i}_{x_{L}} \pi^{j}_{x_{R}} = \pi^{A}_{\vec{n}}\l i ,j \r
\end{equation}
which is a pure operator state on $A$. Bipartitioning $AB$ into $A$ and $B$, we write the Schmidt decomposition as
\begin{equation}
     \pi^{AB}_{\vec{n}}= \sum_{\substack{   i ,\, j,\, k,\, l  \, \in \lcb 0, 1\rcb\\ \lcb \vec{n}_{A}\, \vec{n}_{B} \vert \, \vec{n}_{A} + \vec{n}_{B} = \vec{n} + \vec{V}\l i, j , k , l \r\rcb } }    \pi^{A}_{\vec{n}_{A}} \l i , j \r \pi^{B}_{\vec{n}_{B}} \l  k , l \r 
     \label{eqn:number_restricted_schmidt}
\end{equation}
where $\vec{V}$ counts the isolated quasiparticles living on the boundary sites which are double counted (it is only nonzero when either or both $i=l=1$ and $j=k=1$). Here, $i$ and $j$ are the sites in $B$ most proximate to $A$ and $k$ and $l$ are the sites in $A$ most proximate to $B$. The above equation demonstrates that number projectors have Hartley entropy of at most ${\rm Log}_{2} \lb 4\, {\rm Min}\l N \l A \r, N\l B \r \r\rb$ in bits since each term is a product across the given bipartition. Once again, this statement also applies to the state mappings of these operators by definition.

Schematically, the marginalization of number projectors with normalization follows straightforwardly (we note that for orthogonal projectors the Hilbert-Schmidt norm and trace are equivalent, so for bitstring-diagonal projectors the normalization is done by simple state counting)
\begin{equation}
    \frac{{\rm Tr}_{B}\lb\pi^{AB}_{\vec{n}} \rb}{2^{-\lv AB\rv}} =\sum_{\vec{n}_{A}} \frac{\# \l \vec{n}_{A} \vert \vec{n}_{AB} \r \#\l \vec{n}_{B} \vert \vec{n}_{AB}\r}{\# \l \vec{n}_{AB} \r} \frac{\pi^{A}_{\vec{n}_{A}}}{2^{\lv A\rv}}.
    \label{eqn:marginalized_projector}
\end{equation}
where in the last line $\vec{n}_{B}\equiv \vec{n}_{AB} - \vec{n}_{A}$. From this schematic formula, we can see the von Neumann entropy is that of the subsystem filling distribution on $A$ or $B$ conditioned on global filling on $AB$ of $\vec{n}$ up to $O\l 1 \r$ corrections from the boundary indices.

\subsection{Off-Diagonal Operators}
\label{sect:od_analytics}
For off-diagonal operators in Rule 54, the dephasing argument has some distinctions from the diagonal case. We start with the operator state
\begin{equation}
    \lv\pX_{n} \ket =  \sum_{k,p} \pX_{n}\lv \phi^{k,p} \otimes \tilde{\phi}^{k, p}\ket . 
    \label{eqn:x_state}
\end{equation}
The action of $\pX_{n}$ in the orbit-sector basis is difficult to understand. While the quasiparticle number is not severely violated ($\pX_{n}$ can change $n_{L}$ or $n_{R}$ by at most one each), $\pX_{n}$ has bra and ket bitstrings which lie in many different orbital sectors. Nonetheless, the number sector argument can still be applied to understand refocusing, with the sum of Eqn.~\ref{eqn:number_resolution} modified to
\begin{equation}
     \lv O \ket = \sum_{\vec{N}, \vec{M}} \pi_{\vec{N}} \otimes \tilde{\pi}_{\vec{M}} \ \lv  O \ket
\end{equation}
to account for both the bra and ket occupations.

The refocusing argument still applies because it is a property of any bitstring under Rule 54 dynamics: given any bitstring starting with total occupation $\vec{N}$ and marked quasiparticle pairs (in both the original and dual spaces) at position $x$, the (dual) marked quasiparticles will show back up again at position $x_{\rm rf}\l\vec{N}\r$ ($\tilde{x}_{\rm rf}\l \vec{M}\r$) and time $t_{\rm rf}\l \vec{N}\r$ ($\tilde{t}_{\rm rf}\l \vec{M}\r$). Then, the bra and ket refocusing will occur at time $t_{\rm rf} = {\rm lcm}\l  t_{\rm rf}\l\vec{N}\r, t_{\rm rf}\l\vec{M}\r \r$, which will scale as $O\l L^{2}\r$, in accordance with the dynamical exponent $z=2$. Logarithmic saturation cannot be established rigorously by the same argument as diagonal operators since the operator needs to be evolved past the self-scattering time.

\section{Details of K-flavor Model}
\label{app:Saddle_point}
To guide the reader and provide intuition, we briefly summarize the calculations below. Our goal is to characterize the entropy of distributions of quasiparticle fillings in the so-called $K$-flavor model described in Section~\ref{sect:gen_loe}. In this model, there are $K$ ``flavors'' of pointlike quasiparticles with pairwise distinct and incommensurate time delays. The quasiparticles are non-dispersive. We care about the phase accumulated by a particle as it traverses the system and thus the quasiparticle fillings, $\vec{N} = \l N_{1} ... N_{K}\r$. Filling $N$ quasiparticles on a lattice of size $L$ with $N\lesssim L$, there are many lattice microstates of each filling, $\vec{N}$, so that the probability distribution $p\l \vec{N}\r$ is nontrivial at infinite temperature. 

We verify typical configurations of the $K$-flavor model (in the sense of parametrically dominating the partition function at infinite temperature) have filling $N\sim L$. We also verify that these configurations have relatively uniform filling per flavor, and our partition function is not dominated by configurations in which $N_{\alpha} \sim L $ for $O(1)$ choices of $\alpha$ but by configurations in which $N_{\alpha}\sim O(1)$ for all choices of $\alpha$. This leads to a saddle point approximation, in which we Poissonize the fillings in each flavor, making the problem analogous to the multiplicity of an Einstein solid on $K$ sites. The Shannon entropy computation is then tractable and yields and an extensive entropy.

As a first step, we would like to show that the K-flavor model has an entropy dominated by the filling $N=L$. First, we calculate free energy (here defined simply as the log of the partition function) at infinite temperature and fixed filling $f\l N \r \equiv {\rm Log}\lb \left. \mc{Z} \right|_{N}\rb$ in the limit of $K\to \infty$ and $L\to \infty$ with $K/L$ held constant. In this limit one can verify
\begin{eqnarray}
    f\l N \r  &=& \l N - KL\r {\rm Log} \lb 1 - N/KL\rb \nonumber \\
    && \ \  + N\,  {\rm Log}\lb KL/N \rb \nonumber \\
    &\approx & N\,  {\rm Log}\lb KL/N \rb + N - N^{2}/KL
    \label{eqn:free_energy}
\end{eqnarray}
From here, one can calculate the free energy difference between $N > n $, as
\begin{eqnarray}
    f\l N\r - f\l n\r &=& \l N - n \r {\rm Log}\lb KL/N\rb + N - n  \nonumber \\
    && +\frac{\l n^{2} - N^{2}\r}{KL} - n\,  {\rm Log}\lb N/n \rb.
\end{eqnarray}
When $N$ is greater than $n$ this difference diverges sufficiently quickly in the relevant limit. Thus we may restrict our filling to $N = L$ without loss of generality.

For the K-flavor model there are two broad possibilities for the saddle point contribution to the partition function, as detailed in the main text: (i) $O\l 1 \r$ filling in each flavor (ii) $O\l L \r$ filling in $O\l 1 \r$ flavors. Though, intuitionally, it is expected that the filling be as evenly distributed as possible to maximize entropy the non-trivial multiplicity function means that a more careful approach is justified. We estimate the contributions to the partition function at infinite temperature at fixed filling $N=L$,
\begin{eqnarray}
    &&\mc{Z}_{({\rm i })} \sim \lb \binom{L}{L/K}\rb^{K}  \\
    \mc{Z}_{({\rm ii})} &\sim & {\rm Poly}\l K \r {\rm Poly} \l L \r \lb\binom{L}{O\l L \r}\rb^{\eta} \nonumber
\end{eqnarray}
where $\eta$ is $O\l 1\r $. Then, the corresponding free energies scale as $f_{(i)} \sim L\,  {\rm Log}\lb K \rb$ and $f_{(ii)}\sim L$ to leading order, so we see that (i) dominates (ii) parametrically as $L$ and $K$ are taken sufficiently large with fixed ratio. Additionally, we see that (i) reproduces the leading free energy of the overall partition function. Thus the dominant contribution comes from states with low filling in many different flavors, which in turn allows us to treat each flavor's count as Poissonian.

Now we would like to calculate the entropy of the categorical distribution in the saddle point approximation: this calculation is just like the multiplicity of an Einstein solid with filling $N=L$ and system size $K$. The number of microstates is simply 
\begin{equation}
    \# \ {\rm Microstates} \  =  \ \binom{N + K -1}{ K - 1 }
\end{equation}
and thus the free energy is 
\begin{equation}
\end{equation}
where we remind the reader of the definition $N/K \equiv \rho$. Thus, we find that the entropy scales extensively.

\color{black}

\section{Intermediate Times and Saturation Timescales}
\label{app:early_time}

\subsection{Rule 54: Intermediate Times}

\label{sect:diag_timescales}
\begin{figure}[h]
    \centering
    \includegraphics[width = 0.75\linewidth]{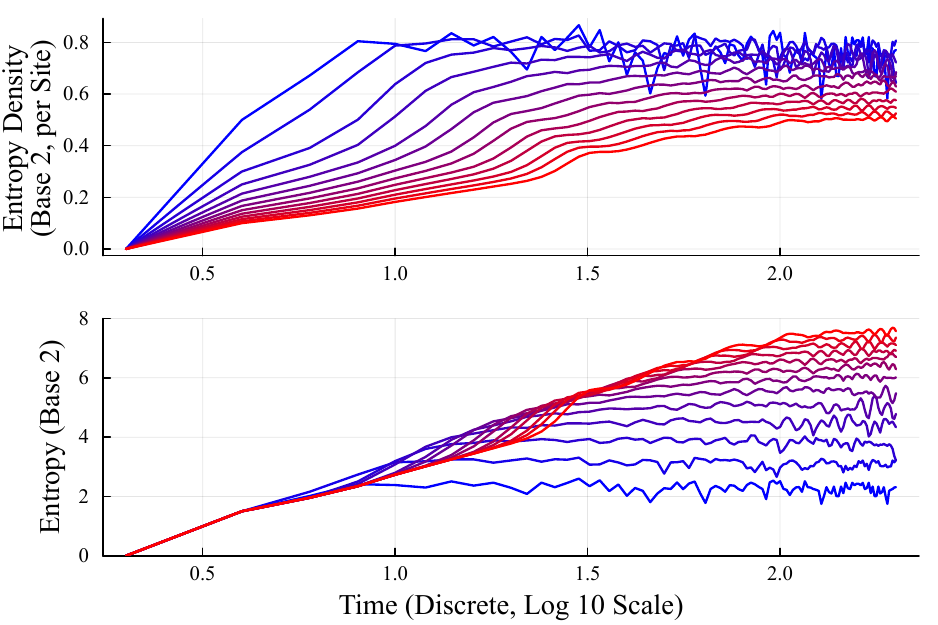}
    \caption{LOE of single-site $\pZ$ gate at early time with half-space bipartition in system sizes of $L=6$ (cold) to  $L=30$ (hot). (Top) LOE entropy density in bits per site. (Bottom) Operator entropy as a function of time.}
    \label{fig:diag_et}
\end{figure}

For diagonal operators, the growth is essentially logarithmic as shown in Fig. \ref{fig:diag_et}.
For early times, we see a uniform in system size logarithmic growth, followed by a period of accelerated growth from $L/2$ to $L$-- see Fig.~\ref{fig:diag_fss}). We note that up to this timescale, there is no entropy growth at the second bipartition interface and the accelerated region is largely accounted for by the second bipartition beginning to contribute before the first is exhausted. It is this effect that is responsible for smaller system sizes initially overtaking larger system sizes-- since the timescale for the second biparition to enter is $O\l L\r$, the smaller system sizes will see its contribution earlier.  After the first interface is saturated, the LOE returns to simple logarithmic growth with fluctuations up to its saturation time.

\label{sect:off_diag_timescales}
For the off-diagonal operators, we can compare directly to the logarithmic bound proposed in \cite{Alba_2021} and find that it is weakly broken by fluctuations after operator self-scattering. However, the qualitative features are robust to the addition of periodic boundary conditions on finite size systems. We note that in a periodic system we must have two bipartition interfaces, so the bounds must trivially be doubled as compared to those discussed in \cite{Alba_2019, Alba_2021}.
\begin{figure}[h]
    \centering
    \includegraphics[width =0.75 \linewidth]{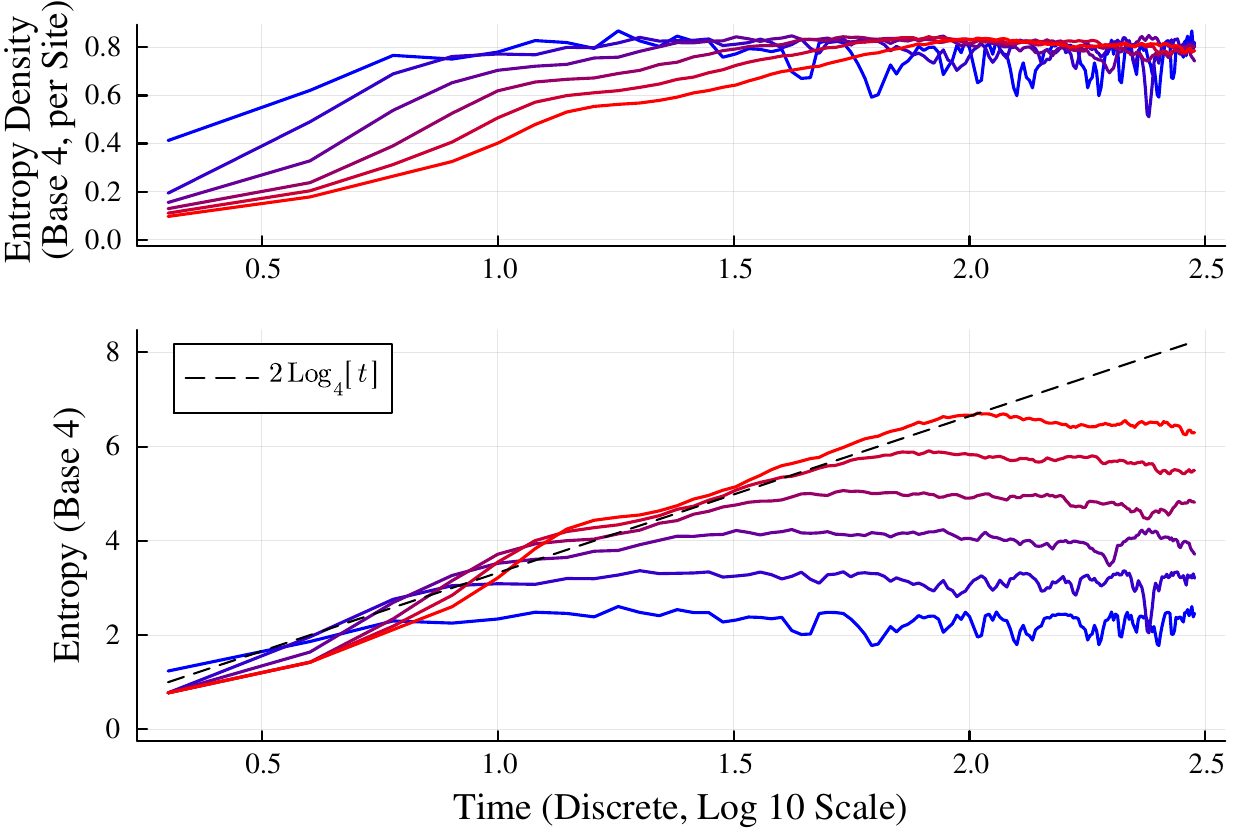}
    \caption{LOE of single-site $\pX$ gate at early time with half-space bipartition in system sizes of $L=6$ (cold) to  $L=16$ (hot). (Top) LOE entropy density in base 4 per site. (Bottom) Operator entropy as a function of time. Logarithmic upper bound as determined in \cite{Alba_2021} (doubled to account for two bipartitions), ${\rm Log}_{4}\lb t \rb$. }
    \label{fig:OD_ET_HS}
\end{figure}
A, perhaps surprising, feature of the data shown in Fig.~\ref{fig:OD_ET_HS} is that the hydrodynamic fluctuations drive the entropy growth to initially overshoot its saturation value. An interpretation for this overshoot in terms of the refocusing picture is not apparent.

\label{sect:Heisenberg_ET}
\subsection{Heisenberg Circuit: Intermediate Times}
\label{sect:log_bounds}
To determine the behaviour of LOE in the Heisenberg model with PBC at short times and compare to the logarithmic bounds given in \cite{Alba_2019,Alba_2021}, we will make use of the integrable trotterization \cite{Vanicat_Trott} of the Heisenberg model, which we will refer to as the Floquet Heisenberg model. It is a $2$-site, $SU(2)$-invariant brickwork circuit given by the gate
\begin{equation}
    U_{n,m} = \frac{1 + i \lambda \mathcal{P}_{n,m}}{1 + i \lambda}.
    \label{eqn:floquet_Heisenberg}
\end{equation}
For the spin $1/2$ case, $\mc{P}_{n,m} = \frac{1}{2}\l\mbs{\sigma}_{n}\cdot \mbs{\sigma}_{m} + 1\r$, which has eigenvalue of $1$ on the symmetric triplet sector and  $-1$ on the antisymmetric singlet sector. This circuit retains an analogue of a Yang-Baxter equation and is meaningfully Bethe-Ansatz integrable. A significant consequence of its brickwork structure is a strict lightcone. The lack of a local energy conservation and the strict lightcone, however, are the only significant distinctions as compared to the continuous-time Heisenberg model. The strict lightcone is helpful to understand the early time dynamics of LOE growth, since the operator growth is upper bounded at all coupling strengths (in particular, we use $\lambda = 1$).

Instead of the log bound of $\frac{2}{3}\, {\rm Log}_{4}\lb t \rb$ (in base $4$ units) conjectured in \cite{Alba_2021} for certain operators in the Heisenberg model, we compare to ${\rm Log}_{4}\lb t \rb$ per bipartition interface as a general reference (not specific to any bound). 
\begin{figure}[h]
    \centering
    \includegraphics[width =0.75 \linewidth]{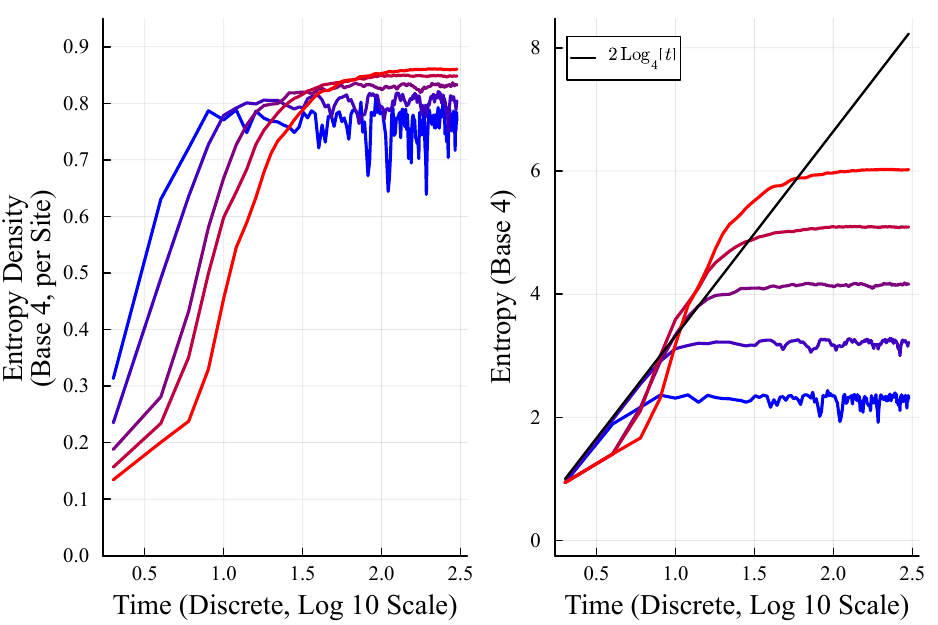}
    \caption{Operator entropy of $\pZ_{1}$ in Floquet Heisenberg for system sizes 6 (cold) to 14 (hot). (Left) LOE density. (Right) LOE compared to $2 \, {\rm Log}_{4}\lb\,  t \,  \rb$. All entropies are in base $4$ units.}
    \label{fig:circuit_Heisenberg}
\end{figure}
The results of Fig. \ref{fig:circuit_Heisenberg} suggest that logarithmic bounds must be broken on periodic boundary conditions. The presence of a logarithmic bound and a volumetric saturation would indicate an exponential saturation timescale (divergent $z$) which, while not fundamentally incompatible, appear challenging to reconcile (\textit{e.g.}, Fig.~\ref{fig:full_timescales_chain_op1} and Fig.~\ref{fig:full_timescales_chain_op3}). We note, however, that these numerical results have little relevance to the case of LOE growth on infinite systems and they should not be directly compared to those of \cite{Alba_2019}.

\subsection{Dynamical Scaling Concepts}

Generally, the timescale analysis can be broken into four cases delineated by the saturation (volume or log law) and growth (logarithmic or algebraic). To organize these scenarios, we can refer to the dynamical scaling relation, 
\begin{equation}
    S \l t \r\sim  L^{\alpha} \, u\l t / L^{z}\r
    \label{eqn:family_vicsek}
\end{equation}
where we define
\begin{equation}
    u\l x \r =
    \begin{cases}
        x^{\beta} & x \ll 1 \\
        1 & x\gg 1 
    \end{cases}
\end{equation}
and the cases interpolating between $x\gg1$  and $x\ll 1$ are left unspecified. We also define $x^{0} = \log{x}$ and $x^{\infty} = \exp\l x \r$. Generally, it is true that $z\beta = \alpha$. However, when $\alpha$ goes to zero (\textit{i.e.}, saturation scales as a log), either $z$ or $\beta$ must go to zero with the other entirely unspecified. Of course, $\alpha>1$ is strictly forbidden by the upper bound of the von Neumann entropy, $S_{\rm vN}^{A} \leq \log{{\rm dim}\l\mc{H}^{A}\r}$.

\label{app:dynamical_scaling}
\emph{Scaling Collapse}.---To extract the exponents from numerical data, one performs a data collapse. A generally applicable method for performing such a collapse on a dynamical quantity $f\l L, t\r$ is as follows:
\begin{enumerate}
    \item Calculate the late-time averaged value of $f\l L,t \r$, as a function of system size: $$f_{\infty}\l L\r =\lim_{T, \,\delta t \to \infty}\frac{1}{\delta t }\int_{T}^{T+\delta t} f\l L, t \r \ dt. $$ Plotting $\log{f_{\infty}\l L \r}$ against $\log{L}$, if Eqn. (\ref{eqn:family_vicsek}) holds, the slope will converge to $\alpha$ for sufficiently large $\log{L}$. 
    \item With $\alpha$ determined, we can extract $u\l t /L^{z}\r \sim f\l L, t \r/f_{\infty}\l L\r$. Plotting $f\l L, t \r/f_{\infty}\l L\r$ as a function of $t/L^{z}$, it will collapse onto a single, well-defined function for sufficiently large $L$ with the correct choice of $z$.
    \item From here the exponent $\beta$ can be deduced from $\alpha$ and $z$, but it can also be extracted directly from a log-log plot of $u$ as a function of $x=t/L^{z}$.
\end{enumerate}
Since data collapses are not always clean for numerically accessible system sizes, it is is sometimes helpful to make use of the simpler threshold method.

\label{app:threshold_method}
\emph{Threshold Method}.--- One very elementary method to determine $z$, is to examine the scaling of a dynamical quantity as it crosses a threshold. If we have early-time sample set $\lcb \tau_{n}\rcb = \mc{T}$, then we say that the threshold times are defined
\begin{equation}
    \Omega^{O}\l L; \zeta \r \equiv {\rm min} \lcb \tau_{n} \in \mc{T} \, : \, f\l L, t \r / f_{\infty}\l L \r  > \,  \zeta \rcb  .
    \label{eqn:threshold}
\end{equation}
Assuming that $\Omega\l L\r$ demonstrates algebraic behaviour in $L$ for fixed $\zeta$, we can assume scaling form
\begin{equation}
     \Omega^{O}\l L; \zeta \r = L^{z} f\l \zeta \r
\end{equation}
then extract $z$ by plotting $\Omega$ on a log-log scale against $L$ with various $\zeta$ fixed.

\subsection{Rule 54: Saturation Timescales}
We now turn to a scaling analysis of the saturation time. In the case of diagonal operators in Rule 54 we will only use the dynamical scaling method as the operator self-scattering time does not parametrically separate from the saturation time. For both diagonal and off-diagonal operators we find that the saturation times scale like the self-scattering times in $L$: $L$ and $L^{2}$ respectively.

\label{sect:r54_et_timescales_diag}

\begin{figure}[h]
    \centering
    \includegraphics[width = 0.75\linewidth]{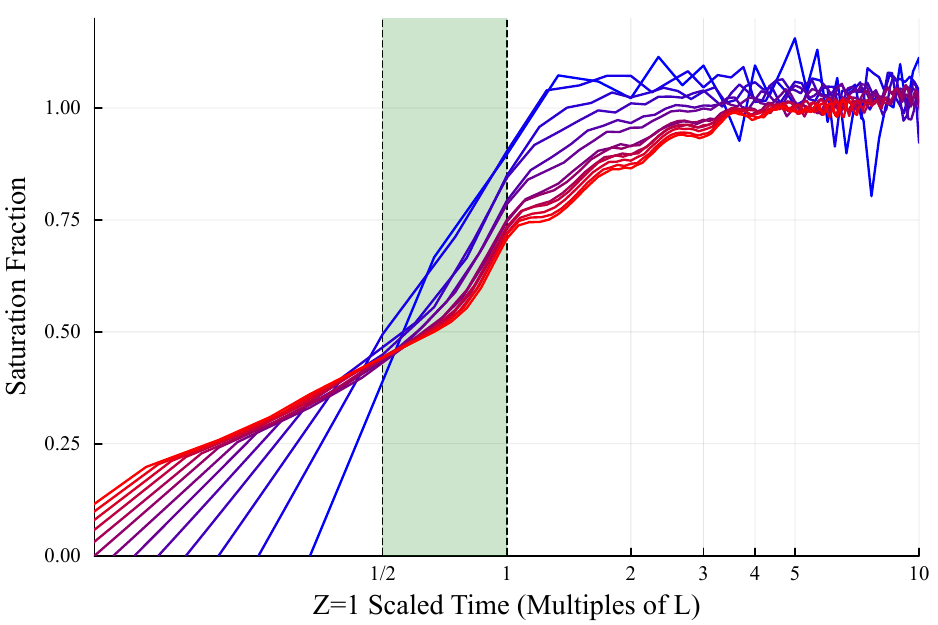}
    \caption{Data Collapse for diagonal operator $\pZ_{1}$ with $z=1$ scaled time, or $t/L$. The $y$-axis is the saturation fraction, or $\svn{O; A }\l t\r/ \svn{O; A}\l\infty\r$. The green highlighted region indicates the operator self-scattering regime, after which strong hydrodynamic fluctuations are evident. The crossing of the x-axis occurring at different values is a sublattice parity effect, whereby the operator has zero entanglement after the first timestep (thus we see the first data point is zero entropy density at $1/L$).}
    \label{fig:diag_fss}
\end{figure}

Using $z=1 $ scaled time (namely, $t/L$) data collapse is shown in Fig. \ref{fig:diag_fss}, which is derived from the same data as Fig. \ref{fig:diag_et}. The exponents are consistent with a logarithmic growth up to saturation scale $2\, {\rm Log}_{2}\lb L \rb$, implying the prefactor of the logarithmic growth is also upper bounded by $2$ (in base $2$ rather than base $4$ units), which is consistent with \cite{Alba_2021} when both bipartition interfaces are accounted for. In terms of the dynamical exponents $\alpha$, $\beta$, and $z$, this is $\alpha = \beta = 0 $ with $z=1$. Since the operator self-scattering time and the saturation time both scale with $L$, the operator self-scattering does not interfere with the data collapse (in contrast to the off-diagonal case).

\label{sect:r54_et_timescales_off_diag}

\begin{figure}[h]
    \centering
    \includegraphics[width =0.75 \linewidth]{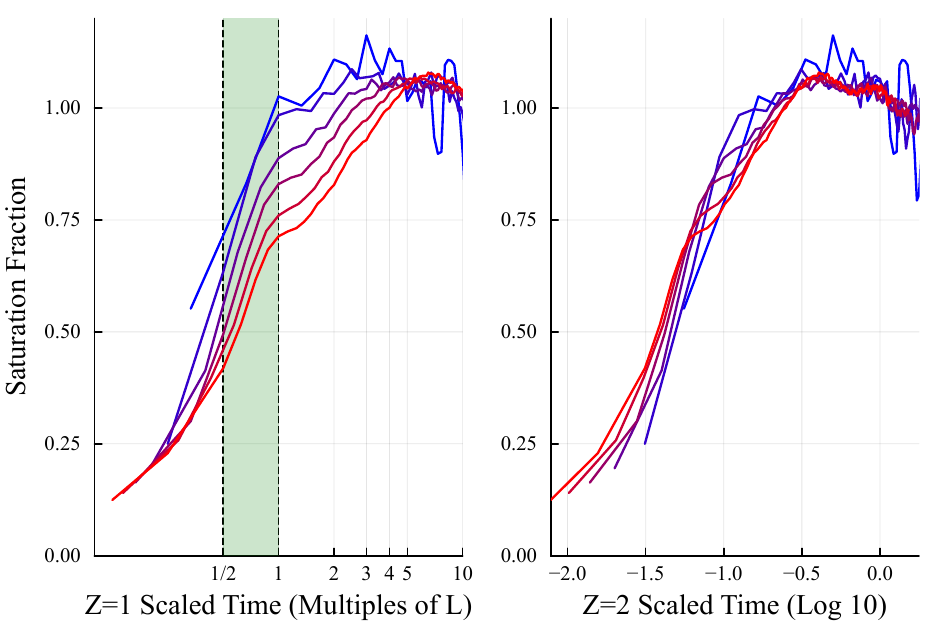}
    \caption{Scaled data for the off-diagonal operator $\pX_{1}$. The $x$-axis is scaled time (\textit{i.e.}, $t/L^z$) and the $y$-axis is saturation fraction, or $\svn{O; A }\l t\r/ \svn{O; A}\l\infty\r$. (Left) $z=1$ scaled time, showing the operator self-scattering regime highlighted in green. (Right) $z=2$ scaled time showing the long-time data collapse past the self scattering regime.}
    \label{fig:OD_fss}
\end{figure}

\begin{figure}[h!]
    \centering
    \includegraphics[width = 0.75\linewidth]{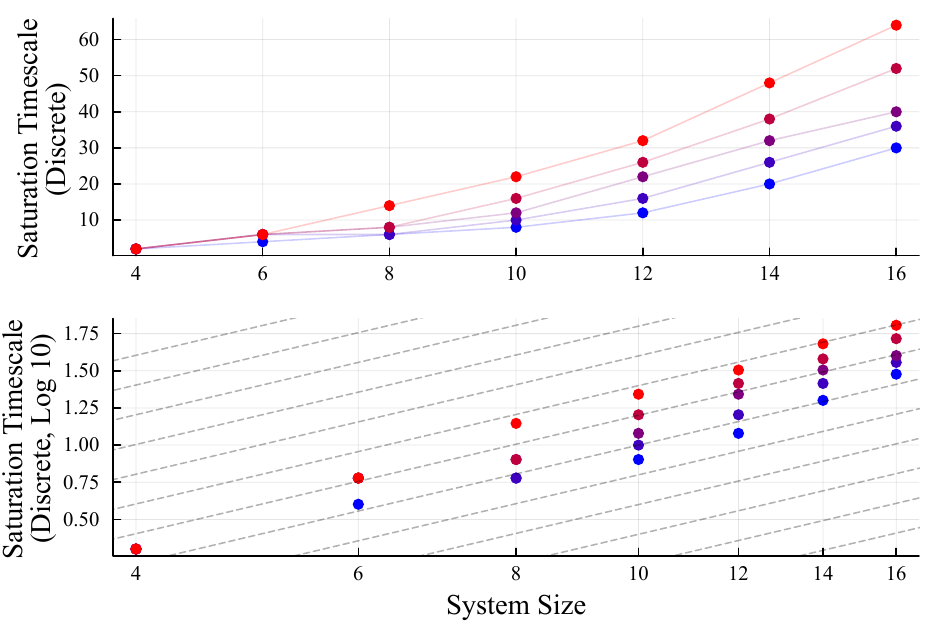}
    \caption{Timescales of half-space bipartition LOE with initial operator $\pX_{1}$. Threshold method with $\zeta$ (fraction of saturated value for threshold) going from $0.7$ (cold) to $0.95$ (hot) in $0.05$ increments. (Top) Direct threshold times on linear scale. (Bottom) Log-log plot of threshold times. Dashed diagonals are ``diffusion lines,'' which have slope $2$ (indicating the dynamical exponent for diffusion).}
    \label{fig:full_hs_timescales}
\end{figure}
The direct finite size scaling method for off-diagonal operators is not as straightforward as for diagonal operators since the operator self-scattering time (which is $O\l L \r$) has not fully separated from the saturation time (which is $O\l L^{2}\r$) for the system sizes considered. The effect of the self-scattering is, however, limited and we can still extract a relatively clear dynamical exponent of $z=2$ using the data collapse presented in the right panel of Fig. \ref{fig:OD_fss} and the elementary threshold method of Fig. \ref{fig:full_hs_timescales}.

\subsection{Heisenberg Chain: Saturation Timescales}
\begin{figure}[h!]
    \centering
    \includegraphics[width =0.75 \linewidth ]{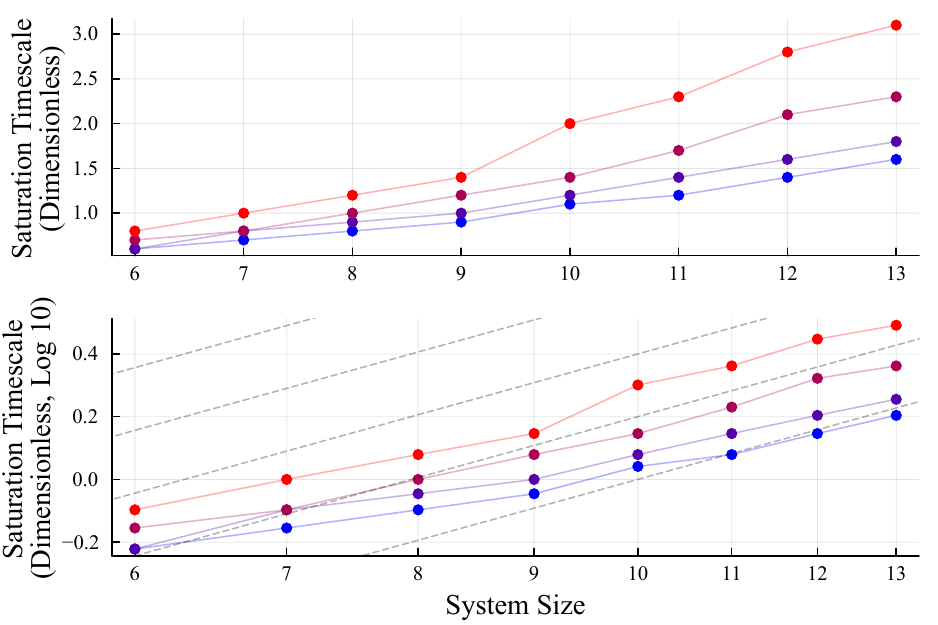}
    \caption{Timescales of operator $\pZ_{1}$. Threshold method with $\zeta$ (fraction of saturated value for threshold) going from $0.6$ (cold) to $0.9$ (hot) in $0.1$ increments. (Top) Direct threshold times on linear scale. (Bottom) Log-log plot of threshold times. Dashed diagonals are diffusion lines ($z=2$).}
    \label{fig:full_timescales_chain_op1}
\end{figure}

\begin{figure}[h!]
    \centering
    \includegraphics[width =0.75 \linewidth ]{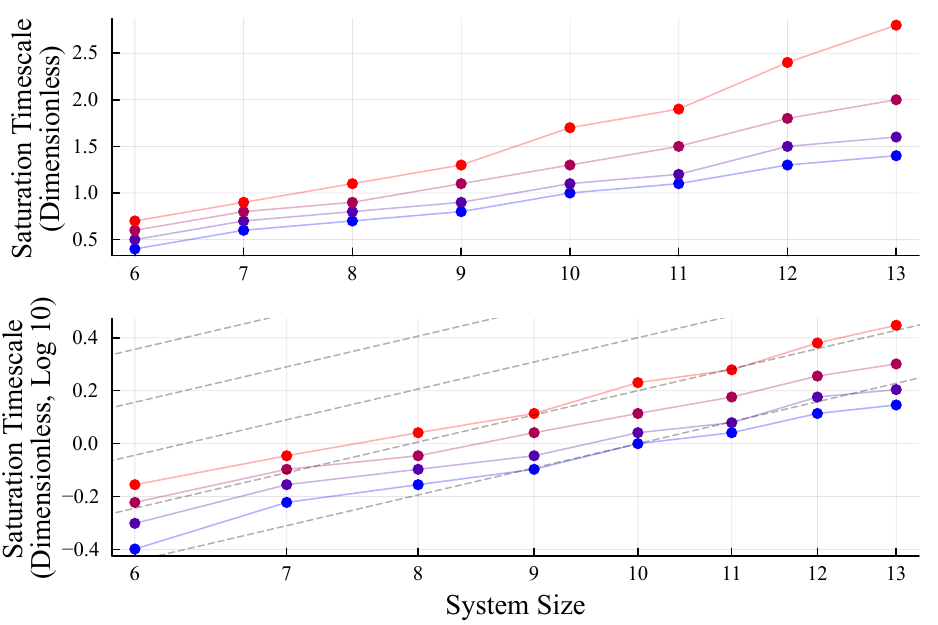}
    \caption{Timescales of operator $\pZ_{1}\pZ_{2}$. Threshold method with $\zeta$ (fraction of saturated value for threshold) going from $0.6$ (cold) to $0.9$ (hot) in $0.1$ increments. (Top) Direct threshold times on linear scale. (Bottom) Log-log plot of threshold times. Dashed diagonals are diffusion lines ($z=2$).}
    \label{fig:full_timescales_chain_op3}
\end{figure}

The saturation timescale in the Heisenberg model presents a substantial challenge for multiple reasons. First, one cannot make use of the symmetry resolution scheme used for the saturation values. Thus, we are limited to small system sizes, and finite size effects are pronounced. Second, the data collapse is ambiguous, showing stronger agreement in some regions than others depending on the chosen exponent. Due to this difficulty, we only show the data processed via the threshold method in Figs. \ref{fig:full_timescales_chain_op1} and \ref{fig:full_timescales_chain_op3}, which show very similar results. The results are inconclusive, showing some agreement with $z=2$.

\section{\titlemath{U(1)} Subgroup and Numerical Methods}
\label{app:symmetry_resolution}
\subsection{Symmetry Principles}
We can make use of the $U(1)$ subgroup of the Heisenberg model by breaking down our problem into different subspaces labeled by eigenvalues of $ \sum_{x \in \mathscr{L}} Z_{x}$. Let $\mc{Q}_{n}$ be the subspaces of definite $U(1)$ charge, labeled by $n$. Thus, if we choose an operator which is uncharged (and therefore also block diagonal), we can write its dynamics as
\begin{equation}
    O\l t \r = \bigoplus_{n} e^{- i \left. H \right|_{\mc{Q}_{n}} t }\left. O \right|_{\mc{Q}_{n}}e^{i \left. H \right|_{\mc{Q}_{n}} t }.
\end{equation}
Ideally, we would work individually in different symmetry sectors to determine the time dependent entropy of an operator which is the direct sum of restrictions in multiple symmetry sectors. A priori, doing this is challenging because the restriction to a symmetry sector breaks the local tensor factorization of the full Hilbert space. However, there is a way to restore this tensor factorization, which we detail in the next section.

\subsection{Symmetry Resolved Entropy}
Our ultimate goal will be to determine the operator entropy of such a state by working only in sectors, but first we will work this procedure out for states. Any state in $\mc{H}$ can be written as a sum across states belonging to each charge sector, with coefficients that remain invariant under time evolution
\begin{equation}
    \lv\Psi \l t\r\ket = \bigoplus_{r}C_{r} \lv \psi\l t\r\in \mc{Q}_{r}\ket \equiv   \bigoplus_{r}C_{r} \lv \mc{Q}_{r}\l t \r\ket,
\end{equation}
where we have introduced shorthand notation in the last equality. It helps to make the additional simplification of assuming the state is only in a single sector, $\mc{Q}$, and therefore we can write it as
\begin{equation}
    \lv \mc{Q} \ket = \bigoplus_{\lcb\left.  \mc{Q}_{\rm A}^{(n)}, \mc{Q}_{\rm B}^{(n)} \right| \mc{Q}_{\rm A}^{(n)}+ \mc{Q}_{\rm B}^{(n)} = \mc{Q} \rcb}  D_{n} \ \lv \mc{Q}_{\rm A}^{(n)} ,\, \mc{Q}_{\rm B}^{(n)} \ket 
    \label{eqn:charge_partition}
\end{equation}
\vspace{20pt}
where $\mathscr{L} = A\cup B$ is an arbitrary but spatially local bipartition. Note that this does not mean that $\lv \mc{Q}_{\rm A}^{(n)} ,\, \mc{Q}_{\rm B}^{(n)} \ket$ is itself $AB$ tensor-factorizable. But we can Schmidt decompose as
\begin{equation}
    \lv \mc{Q}_{\rm A}^{(n)} ,\, \mc{Q}_{\rm B}^{(n)} \ket = \sum_{k}\sqrt{F_{k}^{\ n }} \lv \psi^{\mc{Q}_{A}^{(n)}}_{k} \otimes \psi^{O_{B}^{(n)}}_{k} \ket
    \label{eqn:schmidt_sector}
\end{equation}
with \={\lcb\psi^{\mc{Q}^{n}_{A/B}}_{k}\rcb\in  \mc{Q}_{A/B}^{(n)}}. Then the entropy of a state living in a single \={\mc{Q}_{n}} is simply
\begin{equation}
    \svn{O;\,A} = -\sum_{n,k} \lv  D_{n} \rv^{2} F_{k}^{ \ n } \log{\lv D_{n} \rv^{2} F_{k}^{\ n}}.
    \label{eqn:blocksvn}
\end{equation}
    Now, let us verify this decomposition. We begin in a single symmetry sector, and then we decompose as in Eqn. (\ref{eqn:charge_partition}):
    \begin{eqnarray}
        & &{\rm Tr}_{\rm B}\lb \rho^{\rm AB}\rb = {\rm Tr}_{\rm B}\lb \l \bigoplus_{n} D_{n} \lv  \mc{Q}_{\rm A}^{(n)} ,\, \mc{Q}_{\rm B}^{(n)} \ket\r \l \bigoplus_{m} \bar{D}_{m} \bra  \mc{Q}_{\rm A}^{(m)} ,\, \mc{Q}_{\rm B}^{(m)} \rv \r \rb\nonumber \\
        &=& \sum_{\psi \in \mathcal{H}_{\rm B}} \l \bigoplus_{n} \delta_{\psi \in \mc{Q}^{(n)}_{\rm B}} D_{n} \bra \psi \bigg{\vert} \mc{Q}_{\rm A}^{(n)} ,\, \mc{Q}_{\rm B}^{(n)} \ket \r\l \bigoplus_{m} \delta_{\psi \in \mc{Q}_{\rm B}^{(m)}}\bar{D}_{m}\bra  \mc{Q}_{\rm A}^{(m)} ,\, \mc{Q}_{\rm B}^{(m)}  \bigg{\vert}\psi \ket \r.
    \end{eqnarray}
    In the last line, we have defined the following Kronecker-like \={\delta} function
    \begin{equation}
        \delta_{\psi\in \mc{V}} = 
        \begin{cases}
            1 & \psi \in \mc{V} \\
            0 & \psi \notin \mc{V}
        \end{cases}
    \end{equation}
    which allows the following further simplification
    \begin{equation}
        {\rm Tr}_{B}\lb \rho^{\rm AB}\rb  = \bigoplus_{n} \lv D_{n}\rv^{2}\sum_{\psi \in \mc{Q}_{n}} \bra  \psi \lv  \mc{Q}_{\rm A}^{(n)} ,\, \mc{Q}_{\rm B}^{(n)} \ket \bra  \mc{Q}_{\rm A}^{(n)} ,\, \mc{Q}_{\rm B}^{(n)}\rv \psi\ket = \bigoplus_{n} \lv D_{n}\rv^{2} {\rm Tr}_{\mc{Q}_{B}^{(n)}}\lb\left. \rho^{AB}\right|_{\mc{Q}^{(n)}_{\rm A}\otimes\mc{Q}_{\rm B}^{(n)}} \rb
    \end{equation}
    which is the desired result, from which we conclude that Eqn. (\ref{eqn:blocksvn}) is valid.

For operator states, the generalization is straightforward, if detailed. To begin we consider an operator which is \={U\l1\r} uncharged and therefore block diagonal. Thus, we can write \={O} as
\begin{equation}
    O = \bigoplus_{r}C_{r}\left. O\right|_{\mc{Q}_{r}} = \bigoplus_{r}C_{r}\sum_{\psi_{k},\psi_{l} \in \mc{Q}_{r}}\left.O_{k,l}\right|_{\mc{Q}_{n}}\lv \psi_{k} \ket\bra \psi_{l}  \rv.
\end{equation}
Trivially, the operator state mapping becomes
\begin{equation}
    \lv O\ket = \bigoplus_{r}C_{r}\sum_{\psi_{k},\psi_{l} \in \mc{Q}_{r}}\left.O_{k,l}\right|_{\mc{Q}_{r}}\lv \psi_{k},   \tilde{\psi}_{l}  \ket
\end{equation}
and we can simplify by a change of basis to get
\begin{equation}
    \lv O \ket = \bigoplus_{r} C_{r} \lv  \mc{Q}_{r}, \tilde{\mc{Q}}_{r}\ket
\end{equation}
where the \={\lcb\lv  \mc{Q}_{r}, \tilde{\mc{Q}}_{r}\ket \rcb} are one representative state in each \={\mc{Q}_{r}\otimes \tilde{\mc{Q}}_{r}} sector. Now, we can perform a decomposition analogous to Eqn. (\ref{eqn:charge_partition}), but subject to a different constraint. Instead of \={\mc{Q}_{\rm A}^{(n,r)} + \mc{Q}_{\rm B}^{(n,r)} = \mc{Q}_{\rm r}} we require that \={\mc{Q}_{\rm A}^{(n,r)}+ \mc{Q}_{\rm B}^{(n,r)} = \tilde{\mc{Q}}_{\rm A}^{(n,r)}+ \tilde{\mc{Q}}_{\rm B}^{(n,r)} = \mc{Q}^{r}} \footnote{Which also implies $\mc{Q}_{\rm A}^{(n,r)} + \mc{Q}_{\rm B}^{(n,r)} - \tilde{\mc{Q}}_{\rm A}^{(n,r)} - \tilde{\mc{Q}}_{\rm B}^{(n,r)} = 0$, the condition for an operator to not carry charge.}.
    Thus, we can rewrite \={\lv  \mc{Q}_{r}, \tilde{\mc{Q}}_{r}\ket} as 
    \begin{equation}
        \lv  \mc{Q}_{r}, \tilde{\mc{Q}}_{r}\ket = \bigoplus_{\substack{\lcb\left.  \mc{Q}_{\rm A}^{(n)}, \mc{Q}_{\rm B}^{(n)} \right| \mc{Q}_{\rm A}^{(n)}+ \mc{Q}_{\rm B}^{(n)} = \mc{Q}_{r} \rcb \\ \lcb\left.  \tilde{\mc{Q}}_{\rm A}^{(m)}, \tilde{\mc{Q}}_{\rm B}^{(m)} \right| \tilde{\mc{Q}}_{\rm A}^{(m)}+ \tilde{\mc{Q}}_{\rm B}^{(m)} = \mc{Q}_{r} \rcb }} D_{n,m}^{(r)} \lv  \mc{Q}_{\rm A}^{(n)} , \mc{Q}_{\rm B}^{(n)},  \tilde{\mc{Q}}_{\rm A}^{(m)}, \tilde{\mc{Q}}_{\rm B}^{(m)}\ket.
    \end{equation}
    These basis states can be Schmidt decomposed as
    \begin{equation}
        \lv \mc{Q}^{(n)}_{\rm A}, \mc{Q}^{(n)}_{ \rm B}, \tilde{\mc{Q}}^{(m)}_{\rm A}, \tilde{\mc{Q}}^{(m)}_{\rm B} \ket = \sum_{k} \sqrt{F_{k}^{ \ n, m }} \lv \Psi^{\mc{Q}_{\rm A}^{(n)} \otimes \tilde{\mc{Q}}_{\rm A}^{(m)}}_{k} \otimes \Psi^{\mc{Q}_{\rm B}^{(n)} \otimes \tilde{\mc{Q}}_{\rm B}^{(m)}}_{k}  \ket.
    \end{equation}
    We will once again make the simplification of assuming that there is only a single non-zero \={C_{r}} to begin. Then, the reduced density matrix is 
    \begin{eqnarray}
        \rho^{\rm A}_{O} &=& \sum_{s,t} \sum_{k,l} \bra \psi^{\mc{Q}^{(s)}_{\rm B}}_{k}\tilde{\psi}^{\tilde{\mc{Q}}^{(t)}_{\rm B}}_{l}\rv  \l \bigoplus_{n,m}  D_{n,m}  \lv  \mc{Q}_{\rm A}^{(n)} , \mc{Q}_{\rm B}^{(n)},  \tilde{\mc{Q}}_{\rm A}^{(m)}, \tilde{\mc{Q}}_{\rm B}^{(m)}\ket\r\nonumber \\
        & & \times \l \bigoplus_{p,q  } \bar{D}_{p,q } \bra \mc{Q}_{\rm A}^{(p)} , \mc{Q}_{\rm B}^{(p)},  \tilde{\mc{Q}}_{\rm A}^{(q)}, \tilde{\mc{Q}}_{\rm B}^{(q)} \rv\r \lv \psi^{\mc{Q}^{(s)}_{\rm B}}_{k}\tilde{\psi}^{\tilde{\mc{Q}}^{(t)}_{\rm B}}_{l}\ket \nonumber \\
        &=& \sum_{n,m,k,l} \lv D_{n,m} \rv^{2}  \delta_{n,p}\delta_{m,q} \bra \psi^{\mc{Q}^{(n)}_{\rm B}}_{k}\tilde{\psi}^{\tilde{\mc{Q}}^{(m)}_{\rm B}}_{l} \bigg{\vert} \mc{Q}_{\rm A}^{(n)} , \mc{Q}_{\rm B}^{(n)},  \tilde{\mc{Q}}_{\rm A}^{(m)}, \tilde{\mc{Q}}_{\rm B}^{(m)}\ket \nonumber  \\
        & & \times  \bra \mc{Q}_{\rm A}^{(p)} , \mc{Q}_{\rm B}^{(p)},  \tilde{\mc{Q}}_{\rm A}^{(q)}, \tilde{\mc{Q}}_{\rm B}^{(q)} \bigg{\vert} \psi^{\mc{Q}^{(s)}_{\rm B}}_{k}\tilde{\psi}^{\tilde{\mc{Q}}^{(t)}_{\rm B}}_{l}\ket \nonumber \\
        &=& \sum_{n, m, k } \lv D_{n, m }\rv^{2} F_{k}^{ \ n,m} \lv \Psi^{\mc{Q}_{\rm A}^{(n)} \otimes \tilde{\mc{Q}}_{\rm A}^{(m)}}_{k}   \ket \bra \Psi^{\mc{Q}_{\rm A}^{(n)} \otimes \tilde{\mc{Q}}_{\rm A}^{(m)}}_{k} \rv.
    \end{eqnarray}
    The entropy then can be written:
    \begin{equation}
        \svn{O; A } = -\sum_{n,m,k} \lv D_{n,m}\rv^{2} F_{k}^{n,m} {\rm}\,  {\rm Log}_{4}\lb\lv D_{n,m}\rv^{2} F_{k}^{n,m} \rb.
    \end{equation}

This derivation is not as useful for the case of operators which live in multiple sectors. However, the projection seemingly does not matter much at late times in practice (see Fig. \ref{fig:u1_comparison}). However, it may be of use for less coarse metrics and thus could be a fruitful future direction. 

\subsection{\titlemath{U(1)} Projection and Time Regimes}
\label{app:u1_comparison}
\begin{figure}[h]
    \centering
    \includegraphics[width = 0.75\linewidth]{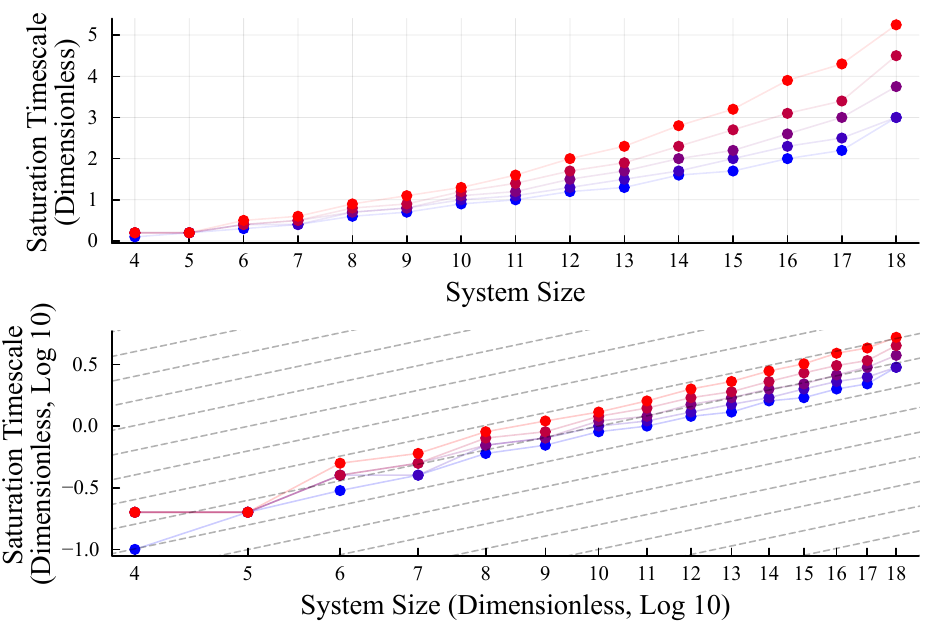}
    \caption{$U(1)$ projected timescales of operator $\npZ_{1}\npZ_{2}$. Threshold method with $\zeta$ going from $0.7$ (cold) to $0.95$ (hot) in $0.05$ increments. (Top) Direct threshold times on linear scale. (Bottom) Log-log plot of threshold times. Dashed diagonals are ``diffusion lines,'' which have slope $2$ (indicating the dynamical exponent for diffusion). System size $L=18$ has a coarser time sample than smaller system sizes, resulting in a slight shift in the data. }
    \label{fig:threshold_zz_u1}
\end{figure}

We note that projecting into a single symmetry sector has significant impact on early time dynamics. If one uses this symmetry projection scheme to extract timescales, it returns tightly diffusive results as shown in Fig. (\ref{fig:threshold_zz_u1}).

\begin{figure}[h!]
    \centering
    \includegraphics[width =0.75 \linewidth ]{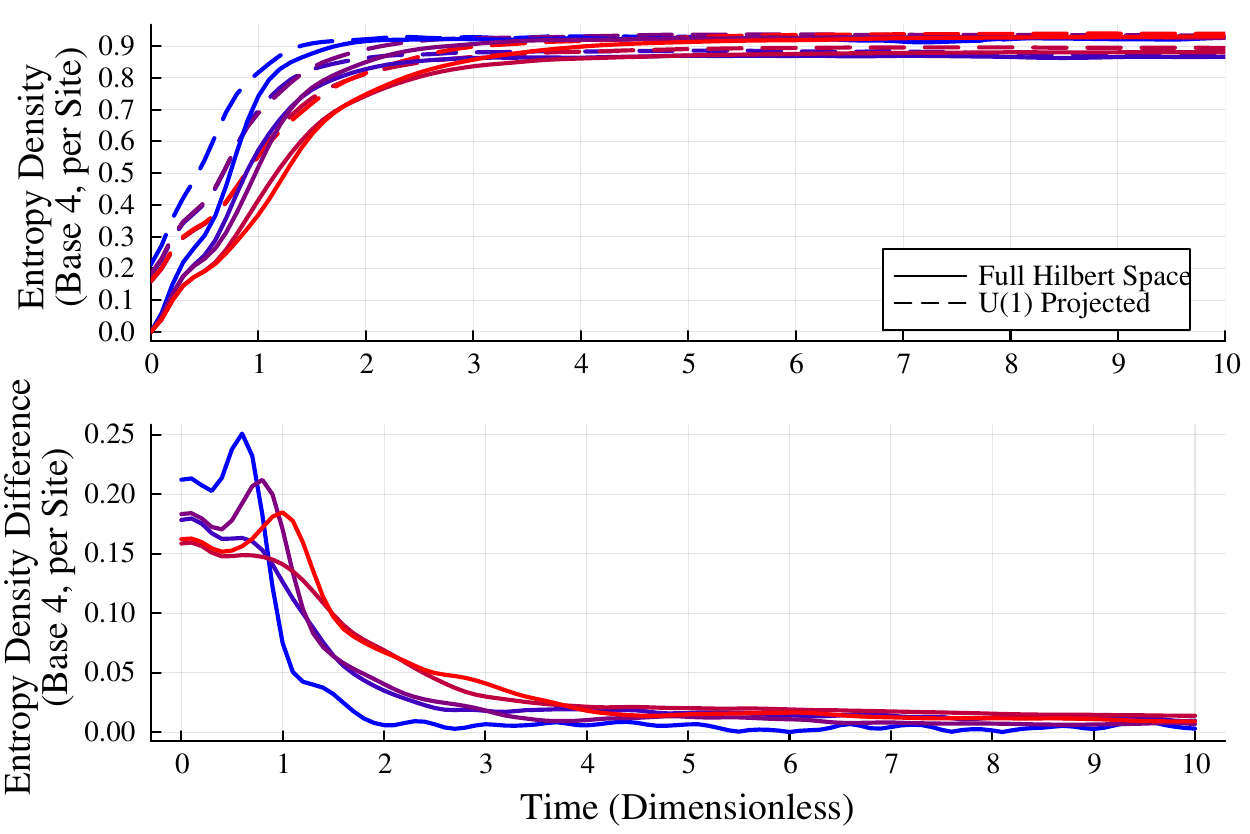}
\caption{Differences between full Hilbert space and $U(1)$ projected LOE (top) and LOE density (bottom) for system sizes 7 (cold) to 13 (hot) with initial operator $\npZ_{1}$.}
    \label{fig:u1_comparison}
\end{figure}

Nonetheless, per Fig. (\ref{fig:u1_comparison}), starting at approximately the saturation timescale the $U(1)$ projection scheme demonstrates very strong agreement with the non-projected results. We expect that at late time, the lack of coherence between different $U(1)$ sectors in the Heisenberg model should render the projection scheme essentially irrelevant to the operator entropy. 

\section{Computational Methods for CA Dynamics}
\label{app:r54_calc}

\subsection{Computational Tools for Diagonal Operator Evolution and Time Averages}
\label{app:diag_tools}
Rather than an explicit implementation of the time evolution operators introduced in Eqn. (\ref{eqn:r54_u_n}), it is vastly more efficient to introduce a dictionary of states and their mappings, leveraging the bitstring representation of computational basis states. If we vectorize the computational basis on $\mc{H}$, which we will denote $\sum_{i}\mathbf{\hat{e}}_{i}\lv\phi^{i} \ket = \vec{\lv \phi\ket}$, then we can apply the time evolution elementwise. If we decompose a state as $\lv \psi \ket = \vec{\psi} \cdot \vec{\lv\phi \ket}$, we can say that $\lv \psi\l t\r\ket = \vec{\psi} \cdot \vert \vec{\phi}\l t \r \rangle$. Now we define computational basis states as $\lv\phi^{i} \ket = \bigotimes_{n=1}^{L}\lv \phi_{n}^{i}\ket$ and introduce the map
\begin{equation}
    \mathscr{F}\lb \lv\phi^{i} \ket\rb = 1+  \sum_{n=0}^{L-1} \phi_{n}^{i}2^{n} ,
\end{equation}
which relates the bitstring and the element of vector basis. Now one can define a time evolution on a state by the following transformation:
\begin{eqnarray}
    &&\vec{\psi} \cdot \vert  \vec{\phi} \l 2 \r \rangle =   \sum_{i} \psi_{i} \lv \phi^{i}\l 2 \r\ket = \sum_{i} \psi_{i} \vert \phi^{\mathscr{F}\lb U_{54} \vert \phi^{i}\rangle\rb} \rangle \nonumber \\
    &\equiv& \sum_{i} \psi_{i}\, \vert \phi^{\sigma\l i\r}\rangle = \sum_{i} \psi_{\sigma^{-1}\l i \r} \vert \phi^{i}\rangle
    \label{eqn:diag_time_evolution}
\end{eqnarray}
where we have used the fact that a sum is invariant under a permutation of its indices in the last step. We can work inductively from the above calculation as follows
\begin{equation}
    \psi_{i}\l t+2 \r =\psi_{\sigma^{-1}\l i \r}\l t\r \qquad \lv \psi\l 2t\r\ket = \sum_{i} \psi_{\sigma^{\l -t\r}\l i\r} \vert \phi^{i}\rangle
\end{equation}
Note that to maintain a uniform $\sigma$ for time evolutions, only the full Floquet cycle operator was used (with both even and odd updates). Computationally, this operation is a simple reshape. To determine late time behaviour, it is most straightforward to exponentiate by squares:
\begin{equation}
    \sigma^{\l 2^{n+1}\r} = \sigma^{\l 2^{n}\r}\l \sigma^{\l 2^{n}\r}\r
\end{equation}

Should the inversion of $\sigma$ prove prohibitive, it can in most cases be forgone or otherwise accounted for simply. In Rule 54 time reversal is equivalent to conjugation by a single lattice site translation since
\begin{eqnarray}
    && TU_{\rm even}U_{\rm odd}T^{-1} U_{\rm even} U_{\rm odd}  \nonumber \\
    &=&TU_{\rm even}T^{-1}T U_{\rm odd}T^{-1} U_{\rm even} U_{\rm odd} \nonumber \\
    &=& U_{\rm odd}U_{\rm even}U_{\rm even} U_{\rm odd} = \mathbb{I}.
    \label{eqn:time_reversal}
\end{eqnarray}
The final identity follows from the fact that if $s_{n}' = s_{n-1}+s_{n} + s_{n+1}+s_{n-1}s_{n+1}\mod{2} $ and $s_{n}'' = s_{n-1}+s_{n}' + s_{n+1}+s_{n-1}s_{n+1}\mod{2} $, then $s_{n}'' = 2s_{n-1}+s_{n} + 2s_{n+1}+2s_{n-1}s_{n+1}\mod{2}  = s_{n} $ so that $U_{\rm even}^{2} = U_{odd}^{2} = \mathbb{I}$. Thus it is sufficient to simply conjugate by a translation operator to invert.

\subsection{Time Evolution of Off-Diagonal Operators}
\label{app:off_diag_tools}
For diagonal operators in Rule 54, the time evolution is extremely simply defined by Eqn. (\ref{eqn:diag_time_evolution}). However, off-diagonal operators are somewhat more complicated. We begin with a local operator that has decomposition into the computational basis states
\begin{equation}
    O = \sum_{i, j } O_{i\,  j } \lv\phi^{i} \ket\bra \phi^{j}\rv  .
\end{equation}
The operator cannot be mapped to a non-doubled state; however, the sparsity of local operators means that a sparse matrix representation reduces the memory scaling in the computational basis from $4^{L}$ to at worst $3\cdot4^{\lv\mathscr{U}\rv + \frac{1}{2}\lv\overline{\mathscr{U}}  \rv}$ where $\mathscr{U}$ is the lattice subset on which the operator is initially supported and $\overline{\mathscr{U}} = \mathscr{L} \setminus \mathscr{U}$ is the complement. Since Rule 54 dynamics acts on computational basis states as a permutation, no new matrix elements are made and this scaling behaviour remains constant throughout evolution, in contrast to typical sparse representations.

Leveraging this sparse representation for time evolution is very simple:
\begin{equation}
    O\l t \r = \sum_{i, j} O_{ij} \lv \phi^{\sigma\l i \r}\ket\bra \phi^{\sigma\l j\r}\rv
    \label{eqn:od_time_evolution}
\end{equation}
which can be implemented as a transformation on the vectors which store the indices of the non-zero values. The reshape can be pre-computed and the reduced density matrix derived in a sparse representation, but extracting the entropy requires a return to a dense representation.

\subsection{Comment on \titlemath{{\rm L} \div 3} \textit{v.s.} \titlemath{{\rm L}\div 4 } Partitions for Off-Diagonal Operators}
\label{app:memory}

Upon viewing Fig. \ref{fig:off_diagonal}, a question is apparent: why are larger system sizes displayed for the ${\rm L}\div 3$ partition than the ${\rm L}\div 4$ partition?

For the system sizes examined, evolving the operator itself is nearly trivial, but extracting the entropies is non-trivial. In our simulations, system sizes less than $16$ are computed by reshaping the sparse matrix $O\l t \r$ so that its indices within $\tilde{\mc{H}}^{A}$ and $\mc{H}^{B}$ are transposed (\textit{i.e.}, $\mc{H}_{A} \otimes \tilde{\mc{H}}^{A}$ are the row indices and $\mc{H}^{B} \otimes \tilde{\mc{H}}^{B}$ are the column indices). The reshaped operator is then converted to a dense matrix and its singular values are extracted. However, for system sizes larger than $16$, it was necessary to leave the partially reshaped operator in sparse form and then compute the 
operator density matrix by contracting the indices for $\mc{H}^{B}\otimes \tilde{\mc{H}}^{B}$. Then the eigenvalues can be calculated directly. The memory cost of storing the operator density matrix as a dense matrix is $4^{\lv A\rv}$ rather than the $4^{\lv\mathscr{L}\rv}$ of storing the partially reshaped operator as a dense matrix. The cost of this procedure is that a new computational bottleneck arises: the matrix multiplication to trace the subsystem $B = \bar{A}$. For partition ${\rm L}\div 3$ this matrix multiplication is tenable up to and including system sizes $\lv \mathscr{L}\rv = 24$, whereas for ${\rm L}\div 4$ it becomes prohibitively difficult at $\lv \mathscr{L} \rv =22$.

\end{appendix}





\bibliography{scipost_bib.bib}

\end{document}